\documentclass[12pt]{article}
\pdfoutput=1
\usepackage[margin=3cm]{geometry}
\usepackage[utf8]{inputenc}
\usepackage{amsmath}
\usepackage{amssymb}
\usepackage{authblk} % for authors
\usepackage{xcolor}
\usepackage{physics}
\usepackage{graphicx} % for graphics in pdf_tex figures
\usepackage{slashed}
\usepackage{hyperref}
\usepackage{varioref}       
\usepackage{cleveref}     

\crefname{table}{table}{tables}
\Crefname{table}{Table}{Tables}
\crefname{figure}{figure}{figures}
\Crefname{figure}{Figure}{Figures}

\definecolor{tealblue}{rgb}{0.21, 0.56, 0.63}
\hypersetup{colorlinks=true,allcolors = tealblue,linktocpage=true}

\usepackage{pgf} % for pgf pics, might have to remove on publication if pgfs have to be prerendered

\usepackage[
        backend=bibtex,
        sorting=none,
        style=phys,
        eprint=true,
        doi=false,
        biblabel=brackets
        ]{biblatex}

\bibliography{dm_eft.bib}

\newcommand{\commie}[1]{}

\newcommand\mperiod[1][\rlap]{#1{\ .}}	%punctuation
\newcommand\mcomma[1][\rlap]{#1{\ ,}}

\newcommand{\ud}{\mathrm{d}}

\renewcommand{\d}{\partial}

\numberwithin{equation}{section}
\numberwithin{table}{section}

\newenvironment{eqaed}
    {\begin{equation}
    \begin{aligned}
    }
    { 
    \end{aligned}
    \end{equation}
    \ignorespacesafterend
    }

\begin{document}

\title{Revisiting Dudas-Mourad compactifications}

\author{Ivano Basile\thanks{ivano.basile@umons.ac.be}\;}
\author{Sylvain Thomée\thanks{sylvain.thomee@student.umons.ac.be}}
\affil{\emph{Service de Physique de l'Univers, Champs et Gravitation}\\\emph{Universit\'{e} de Mons - UMONS, Place du Parc 20, 7000 Mons, Belgium}}
\author{Salvatore Raucci\thanks{salvatore.raucci@sns.it}}
\affil{\emph{Scuola Normale Superiore and INFN}\\\emph{Piazza dei Cavalieri 7, 56126, Pisa, Italy}}

\maketitle

\begin{abstract}
    
    Superstring theories in ten dimensions allow spacetime supersymmetry breaking at the string scale at the expense of controlled Minkowski backgrounds. The next-to-maximally symmetric backgrounds, found by Dudas and Mourad, involve a warped compactification on an interval associated with codimension-one defects. We generalize these solutions by varying the effective field theory parameters, and we discuss the dimensional reduction on the interval. In particular, we show that scalars and form fields decouple in a certain range of dimensions, yielding Einstein-Yang-Mills theory. Moreover, we find that the breakdown of this effective description due to light Kaluza-Klein modes reflects the swampland distance conjecture, supporting the consistency of the picture at least qualitatively.

\end{abstract}

\newpage
\tableofcontents
\newpage

\section{Introduction}\label{sec:introduction}

Any attempt to build realistic models in string phenomenology faces at least two main challenges. The first is achieving scale separation, or more generally, settings in which low-energy observers see physics in four spacetime dimensions without moduli. The second is breaking supersymmetry while also obtaining a sufficiently long-lived universe. 

Recent progress on the first issue indicates that scale separation, at least in Anti-de Sitter (flux) compactifications, is very difficult to achieve, and existing proposals are currently under detailed scrutiny~\cite{DeWolfe:2005uu, Tsimpis:2012tu, Gautason:2015tig, Lust:2019zwm, Buratti:2020kda, Lust:2020npd, Junghans:2020acz, Marchesano:2020qvg, Marchesano:2020uqz, DeLuca:2021mcj, DeLuca:2021ojx, Cribiori:2021djm, Emelin:2022cac, Andriot:2022yyj, Cribiori:2022trc, Apers:2022zjx, Tsimpis:2022orc, VanHemelryck:2022ynr}. In particular, the existence of scale-separated compactifications appears to be in tension with criteria arising from the swampland program~\cite{Vafa:2005ui} (see~\cite{Palti:2019pca, vanBeest:2021lhn, Grana:2021zvf} for reviews).

The second issue is also quite severe, since breaking supersymmetry typically entails instabilities (see \emph{e.g.}~\cite{Ooguri:2016pdq, Freivogel:2016qwc} in the context of AdS vacua). While perturbative instabilities can be avoided~\cite{Basile:2018irz, Guarino:2020flh}, nonperturbative decay channels still exist~\cite{Antonelli:2019nar, Dibitetto:2020csn, GarciaEtxebarria:2020xsr, Bomans:2021ara} (see however~\cite{Guarino:2021hrc, Giambrone:2021wsm, Guarino:2022tlw}). The ultimate fate of non-supersymmetric vacua in string theory remains an outstanding open problem. In particular, it is unclear whether the models survive in a stringy regime, potentially described by perturbative (S-)dual frames~\cite{Blum:1997cs, Blum:1997gw, Dienes:1998wu, Basile:2022zee}, decay in supersymmetric states~\cite{Angelantonj:2007ts}, or are completely inconsistent.

This puzzling state of affairs calls for a deeper understanding of both scale separation and supersymmetry breaking in string theory. To this end, a natural starting point is directly breaking supersymmetry at the string scale. This approach is supported by swampland arguments~\cite{Cribiori:2021gbf, Castellano:2021yye, DallAgata:2021nnr}, according to which it would be inconsistent to break supersymmetry at a parametrically low scale. Moreover, evidence is accumulating to the effect that, at least with a sufficient number of supercharges, consistent effective field theories (EFTs) of supergravity are entirely captured by the string landscape~\cite{Montero:2020icj, Hamada:2021bbz, Tarazi:2021duw, Bedroya:2021fbu}. While this could be due to a ``lamppost'' effect, in light of these results it is natural to turn to supersymmetry-breaking mechanisms that arise at the string scale.

Constructions of this type involve projections of ten-dimensional superstrings, and there are only a handful of options that leave no tachyon in the perturbative spectrum. One is the $\text{SO}(16) \times \text{SO}(16)$ heterotic model of~\cite{AlvarezGaume:1986jb, Dixon:1986iz}, which arises from a projection of the $\text{E}_8 \times \text{E}_8$ model. The other two options are orientifold projections~\cite{Sagnotti:1987tw, Pradisi:1988xd, Horava:1989vt, Horava:1989ga, Bianchi:1990yu, Bianchi:1990tb, Bianchi:1991eu, Sagnotti:1992qw} of type 0B and type IIB strings. The former yields the $\text{U}(32)$ ``type $0'\text{B}$'' model of~\cite{Sagnotti:1995ga, Sagnotti:1996qj}, and retains the absence of supersymmetry of the parent theory. The latter, in contrast, yields the $\text{USp}(32)$ model of Sugimoto~\cite{Sugimoto:1999tx}, whereby supersymmetry is broken spontaneously in the open-string sector. This peculiar phenomenon, dubbed ``brane supersymmetry breaking'' (BSB)~\cite{Antoniadis:1999xk, Angelantonj:1999jh, Aldazabal:1999jr, Angelantonj:1999ms, Coudarchet:2021qwc}, is reflected by the presence of a Goldstino gauge-singlet fermion in the spectrum~\cite{Dudas:2000nv, Pradisi:2001yv, Dudas:2000ff, Dudas:2001wd}. See also~\cite{Mourad:2017rrl, Basile:2021vxh, Mourad:2021lma} for reviews.

A remarkable link among the issues of scale separation and supersymmetry breaking emerged in~\cite{Dudas:2000ff}. The low-energy dynamics of the non-supersymmetric string models that we mentioned above contains an exponential potential for the dilaton, leading to string-scale runaway if unbalanced. While one can stabilize the vacuum with fluxes~\cite{Mourad:2016xbk, Antonelli:2019nar}, at least perturbatively~\cite{Basile:2018irz} in a metastable state~\cite{Antonelli:2019nar}, one can alternatively rely on warping. In this fashion, the dilaton varies along the internal space, while leaving no moduli in the low-energy EFT. Furthermore, the resulting Dudas-Mourad vacua of~\cite{Dudas:2000ff} are perturbatively stable~\cite{Basile:2018irz} and scale-separated, since the Ricci-flat spacetime can have a Minkowski factor, while the internal space turns out to be an interval of finite length.

The interval hosts singularities at its ends, thereby casting doubt on the construction at the global level. However, there are reasons to believe that stringy effects would regularize the geometry: these end-of-the-world ``pinch-off'' singularities are universal in these models~\cite{Antonelli:2019nar}, in the sense that they arise independently of the sources placed in the bulk. As a result, one is tempted to infer that they originate from additional defects that need to be included for the triviality of cobordism classes, as advocated in~\cite{McNamara:2019rup} (see also~\cite{Montero:2020icj, Blumenhagen:2021nmi, Andriot:2022mri, Blumenhagen:2022bvh}). The dynamical nature of the gravitational tadpoles driving this intriguing phenomenon thus appears to be closely related to string-scale supersymmetry breaking, and it has been highlighted in~\cite{Buratti:2021fiv, Buratti:2021yia, Angius:2022aeq, Angius:2022mgh} in relation to (a variation of) the distance conjecture~\cite{Ooguri:2006in}.

In this paper, we revisit the Dudas-Mourad solutions, generalizing them to other values of the EFT parameters, in particular to any spacetime dimension $D$. One motivation for doing so is that various features of the compactification exhibit peculiar patterns varying $D$, and it is instructive to uncover them. This approach is somewhat complementary to that of~\cite{Pelliconi:2021eak}, where several integrable dilaton potentials are explored instead. Moreover, solutions of this type could be potentially relevant to study non-critical string backgrounds. We also discuss in more detail the dimensional reduction over the Dudas-Mourad interval, analyzing the Kaluza-Klein (KK) spectrum of scalar perturbations to address the moduli problem.

This paper is structured as follows. In~\cref{sec:DM_review} we briefly review the Dudas-Mourad solutions arising from non-supersymmetric strings in ten dimensions. These come in two varieties: an orientifold solution, which we discuss in~\cref{sec:DM_orientifold}, and a heterotic solution, which we discuss in~\cref{sec:DM_heterotic}. In~\cref{sec:DM_generalized} we present generalizations of these solutions. In particular, in~\cref{sec:critical_orientifold} and~\cref{sec:critical_heterotic} we focus on two particular values of a parameter which result in simpler expressions, analogous to those of~\cite{Dudas:2000ff}. We also connect our discussion to the cosmological phenomenon of ``climbing'' scalars~\cite{Dudas:2010gi}. In~\cref{sec:freund_rubin_vacua} we focus on the case $D = 10$, relevant for critical strings, where one obtains a 9d Einstein-Yang-Mills theory, which prompts us to address whether further scale separation can be achieved by simple flux compactifications. Although, as expected, this does not work, we nevertheless study the stability of the resulting 9d Freund-Rubin AdS vacua, since they feature some differences with respect to the 10d ones of~\cite{Gubser:2001zr, Mourad:2016xbk, Basile:2018irz} (\emph{e.g.} there is no dynamical dilaton). We do not pinpoint specific unstable modes, but we show that, if they exist, they pertain to the non-abelian gauge sector. We have reasons to believe that all our vacua suffer from instabilities in this sector, the detailed investigation of which is left for future work. Finally, in~\cref{sec:moduli_kk} we assess whether the dimensional reduction on the Dudas-Mourad interval, which has a free parameter classically, produces moduli, studying scalar perturbations and KK masses. In addition to showing the absence of moduli, as a by-product we find that the scalar KK tower satisfies a distance conjecture. We also briefly discuss possible connections between these results and the cobordism (distance) conjecture in~\cref{sec:cobordisms}. We conclude with some closing remarks in~\cref{sec:conclusions}.

\section{A review of Dudas-Mourad vacua}\label{sec:DM_review}

In this section, we review the prototype vacua found by Dudas and Mourad in~\cite{Dudas:2000ff}. These involve only gravity and the dilaton, for which there are two exponential potentials to be considered. 
The first one is relevant for the USp(32) and U(32) orientifold models, where in the string frame there is a ``tadpole potential" proportional to $e^{-\phi}$, indicating the open string origin of supersymmetry breaking.
The second one represents the effective action of the non-supersymmetric heterotic $\text{SO}(16) \times \text{SO}(16)$ model, with the scalar potential emerging from a one-loop effect. In this paper, we focus on the bosonic sector, but one should keep in mind that the solutions we discuss bring along subtleties for fermions~\cite{Mourad:2020cjq, Mourad:2022loy}.

The Einstein-frame effective actions for metric and dilaton that we shall consider are

\begin{eqaed}\label{eq:Dudas_Mourad_action}
    S=\frac{1}{2\alpha'^4}\int d^{10}x \sqrt{-g}\left[R-\frac{1}{2}(\d\phi)^2-2\alpha_{\mbox{\tiny E}} e^{\gamma\phi}\right]\mcomma
\end{eqaed}

\noindent where in $\alpha_{\mbox{\tiny E}}$ we merged the definitions of $\alpha_{\mbox{\tiny E}}$ and $\beta_{\mbox{\tiny E}}$ in the original work.
The tadpole potential forbids ten-dimensional maximally symmetric solutions, and in~\cite{Dudas:2000ff} the authors considered the codimension-one ansatz

\begin{eqaed}\label{eq:DM_ansatz}
    ds^2 &= e^{2A(y)}\eta_{\mu\nu} dx^\mu dx^\nu + e^{2B(y)} dy^2 \mcomma \\
    \phi  &= \phi(y)\mperiod
\end{eqaed}

\noindent For completeness, we mention that the following computations can be extended to the case of any nine-dimensional manifold that is Ricci-flat, instead of taking nine-dimensional Minkowski as in~\cref{eq:DM_ansatz}. Solutions with curved Einstein manifolds are not known in this setup.

In~\cref{eq:DM_ansatz}, one has the freedom to rescale $y$, which can be used to set $B=-\frac{1}{2}\gamma\phi$. The equations of motion now read

\begin{eqaed}\label{eq:DM_eom}
    36(A')^2+8A''+4\gamma A'\phi'+\frac{1}{4}(\phi')^2+\alpha_{\mbox{\tiny E}} & = 0\mcomma \\
    36(A')^2-\frac{1}{4}(\phi')^2+\alpha_{\mbox{\tiny E}} & = 0\mcomma \\
    \phi''+9A'\phi'+\frac{1}{2}\gamma(\phi')^2-2\gamma\alpha_{\mbox{\tiny E}} & = 0\mperiod
\end{eqaed}

\noindent It is then convenient to define 

\begin{eqaed}\label{eq:good_redefinition_in_10D}
    f(y)=\log\left(\sqrt{1+\frac{36 (A')^2}{{\alpha_{\mbox{\tiny E}}}}}+\frac{6A'}{\sqrt{\alpha_{\mbox{\tiny E}}}}\right)\mcomma
\end{eqaed}

\noindent that is well-defined for any sign of $A'$. Written in terms of $f$, the equations of motion are explicitly redundant, which reflects the $y$-reparametrization invariance. Then, $A'$ and $\phi'$ reduce to the simple expressions

\begin{eqaed}\label{eq:A_prime_and_phi_prime_10D}
    A' =\frac{\sqrt{\alpha_{\mbox{\tiny E}}}}{6}\sinh f\mcomma \qquad
    \phi'=\pm 2\sqrt{\alpha_{\mbox{\tiny E}}}\cosh f\mperiod
\end{eqaed}

\subsection{Orientifold case}\label{sec:DM_orientifold}

In the USp(32) and U(32) models, the scalar potential has $\gamma=\frac{3}{2}$, corresponding to $e^{-\phi}$ in the string frame. The only remaining equation of motion in terms of $f$ reads

\begin{eqaed}\label{eq:f_function_10D_orientifold}
    4f'+6\sqrt{\alpha_{\mbox{\tiny E}}}\cosh f \pm 6\sqrt{\alpha_{\mbox{\tiny E}}}\sinh f =0\mcomma
\end{eqaed}

\noindent where the sign choice arises from~\cref{eq:A_prime_and_phi_prime_10D}. The solution is, up to an additive constant,

\begin{eqaed}\label{eq:f_equation_10D_orientifold}
    f=\mp\log\left(\pm\frac{3}{2}\sqrt{\alpha_{\mbox{\tiny E}}} \, y\right)\mperiod 
\end{eqaed}

\noindent From~\cref{eq:A_prime_and_phi_prime_10D} one obtains $A$ and $\phi$, with the sign choice of~\cref{eq:f_equation_10D_orientifold} that cancels after one integration. The metric and the dilaton are (we let $y>0$ for definiteness)

\begin{eqaed}\label{eq:metric_and_dilaton_orientifold}
    ds^2 & = (\sqrt{\alpha_{\mbox{\tiny E}}} \, y)^{\frac{1}{9}}e^{-\frac{1}{8}\alpha_{\mbox{\scalebox{.4}{E}}} y^2}dx_{(9)}^2+ e^{-\frac{3}{2}\varphi_0}(\sqrt{\alpha_{\mbox{\tiny E}}} \, y)^{-1}e^{-\frac{9}{8}\alpha_{\mbox{\scalebox{.4}{E}}} y^2}dy^2 \mcomma \\
    e^{\phi}& = e^{\varphi_0} (\sqrt{\alpha_{\mbox{\tiny E}}} \, y)^{\frac{2}{3}} e^{\frac{3}{4}\alpha_{\mbox{\scalebox{.4}{E}}} y^2}\mperiod
\end{eqaed}

\noindent There are two timelike curvature singularities at $y=0$ and $y\to\infty$, separated by a finite distance. The ``string coupling'' $e^\phi$ vanishes at $y=0$ and diverges as $y\to\infty$.

\subsection{Heterotic case}\label{sec:DM_heterotic}

For the heterotic effective action, the only difference is $\gamma=\frac{5}{2}$, corresponding to no dilaton coupling in the string frame.
Similar considerations lead to

\begin{eqaed}\label{eq:f_equation_10D_heterotic}
    2f'+3\sqrt{\alpha_{\mbox{\tiny E}}}\cosh{f}\pm 5\sqrt{\alpha_{\mbox{\tiny E}}}\sinh{f} =0 \mperiod
\end{eqaed}

\noindent The coefficients of $\sinh$ and $\cosh$ are different, making the functional form of the solution more involved. In fact,~\cref{eq:f_equation_10D_heterotic} has both a trivial solution, that has been discussed in~\cite{Raucci:2022jgw}, and a non-trivial one, that is the focus of this section:

\begin{eqaed}\label{eq:f_solution_heterotic}
    e^f = \pm 2^{\mp 1}\frac{e^{\sqrt{\alpha_{\mbox{\scalebox{.4}{E}}}} \, y} + \varepsilon e^{-\sqrt{\alpha_{\mbox{\scalebox{.4}{E}}}} \, y}}{e^{\sqrt{\alpha_{\mbox{\scalebox{.4}{E}}}} \, y} - \varepsilon e^{-\sqrt{\alpha_{\mbox{\scalebox{.4}{E}}}} \, y}} \mcomma
\end{eqaed}

\noindent with $\varepsilon$ a constant. In $A'$ and $\phi'$, from~\cref{eq:A_prime_and_phi_prime_10D}, the sign choice of~\cref{eq:f_solution_heterotic} results in a sign flip for $\varepsilon$. Hence, that choice can be undone by $\varepsilon\to -\varepsilon$, and without loss of generality, we consider the upper sign in what follows. 

For bulk solutions, $|\varepsilon|$ can be rescaled away and only its sign becomes relevant. Therefore, there are only two non-equivalent (that is, not related by coordinate transformations) solutions, depending on whether $\varepsilon=\pm1$.

If $\varepsilon=1$, for $y>0$ one finds

\begin{eqaed}\label{eq:heterotic_metric_10D}
    ds^2 & = \left(\sinh{\sqrt{\alpha_{\mbox{\tiny E}}} \, y}\right)^{\frac{1}{12}} \left(\cosh{\sqrt{\alpha_{\mbox{\tiny E}}} \, y}\right)^{-\frac{1}{3}}dx_{(9)}^2+ e^{-\frac{5}{2}\varphi_0}\left(\sinh{\sqrt{\alpha_{\mbox{\tiny E}}} \, y}\right)^{-\frac{5}{4}}\left(\cosh{\sqrt{\alpha_{\mbox{\tiny E}}} \, y}\right)^{-5} dy^2\mcomma\\
    e^{\phi} & = e^{\varphi_0} \left(\sinh{\sqrt{\alpha_{\mbox{\tiny E}}} \, y}\right)^{\frac{1}{2}}\left(\cosh{\sqrt{\alpha_{\mbox{\tiny E}}} \, y}\right)^{2}\mperiod
\end{eqaed}

\noindent There are two timelike curvature singularities, at $y=0$ and $\infty$, separated by a finite distance, and $e^\phi$ is again zero at $y=0$ and diverges as $y\to\infty$.

If $\varepsilon=-1$, the resulting solution can be obtained from~\cref{eq:heterotic_metric_10D} after interchanging $\cosh$ with $\sinh$. Curvature singularities are still present at $y=0$ and $\infty$, but the proper $y$-length and the nine-dimensional Planck mass are infinite.

\section{Generalizing the EFT parameters}\label{sec:DM_generalized}

We now generalize the gravitational solution of~\cref{sec:DM_review} to a generic dimension $D$. We consider an exponential dilaton potential with a general coefficient $\gamma$, and study some consequences on gravity, gauge fields and higher forms.

We set our notation by choosing the following action:

\begin{eqaed}\label{eq:DMAction}
    \mathcal{S} = \frac{1}{2 \kappa^2_D} \int \ud^D x \sqrt{-g} \left[ R - \frac{4}{D-2} (\partial \phi)^2 - 2\alpha_{\mbox{\tiny E}} e^{\gamma \phi} \right]\mcomma 
\end{eqaed}

\noindent with $D>2$. The appropriate ansatz is the natural generalization of~\cref{eq:DM_ansatz}, with a codimension-one Ricci-flat $g_{\mu\nu}$. The equations of motion become simpler in the gauge $B=-\frac{1}{2}\gamma\phi$, and read

\begin{eqaed}\label{eq:generalized_DM_eom_D_dimensions}
    (D-2) A^{\prime \prime}+\frac{(D-2)}{2}\gamma \phi' A'+\frac{(D-1)(D-2)}{2} (A')^2+\frac{2}{D-2} (\phi')^2+\alpha_{\mbox{\tiny E}} & = 0 \mcomma\\
    \frac{(D-1)(D-2)}{2} (A')^2-\frac{2}{D-2} (\phi')^2 +\alpha_{E} & = 0 \mcomma \\
    \frac{8}{D-2}\phi''+\frac{8}{D-2}\left[(D-1) A'+\frac{\gamma}{2} \phi' \right] \phi' -2 \gamma \alpha_{E} & = 0 \mperiod
\end{eqaed}

\noindent Only in this section, from now on we work in units $\alpha_{\mbox{\tiny E}}=1$ to simplify the notation. Restoring $\alpha_{\mbox{\tiny E}}$ is straightforward for dimensional reasons.

One can define, similarly to~\cref{eq:good_redefinition_in_10D},

\begin{eqaed}\label{eq:f_definition_general_D}
    f(y)=\log\left[\sqrt{1+\frac{(D-1)(D-2)}{2} (A')^2}+\sqrt{\frac{(D-1)(D-2)}{2}}A'\right]\mcomma
\end{eqaed}

\noindent so that $A'$ and $\phi'$ have simple expressions in terms of $f$. A sign choice arises in $\phi'$, as in~\cref{eq:A_prime_and_phi_prime_10D}.
The equations of motion become equivalent to a single condition for the unknown function $f$. We can further simplify the notation, letting

\begin{eqaed}\label{eq:nice_redefinitions}
    \gamma  = \frac{4\sqrt{D-1}}{D-2}(1+2\delta)\mcomma\qquad
    \xi  = \sqrt{\frac{2 (D-1)}{D-2}} y\mcomma\qquad
    \psi(\xi)  = e^{-f(y(\xi))}\mperiod
\end{eqaed}

\noindent The differential equation for $f$ translates into

\begin{eqaed}\label{eq:remaining_equation_general_D}
        \dot{f}+\cosh{f}\pm (1+2\delta)\sinh{f}=0\mcomma
\end{eqaed}

\noindent where dots stand for derivatives with respect to $\xi$.

A couple of sign choices must be fixed before presenting the solutions. 
First, it is sufficient to work with $\gamma\geq 0$, that is $\delta\geq -\frac{1}{2}$. In fact, negative and positive $\gamma$s differ by a sign flip in~\cref{eq:remaining_equation_general_D}.
Different choices in~\cref{eq:remaining_equation_general_D}, while changing the functional form of solution, do not present additional computational issues: the upper sign leads to 

\begin{eqaed}
    \dot{\psi}=1+\delta-\delta \psi^2\mcomma
\end{eqaed}

\noindent and the lower one leads to the same differential equation for $\lambda(\xi)=-e^{f(\xi)}$.
Therefore, up to interchanging $\cosh$ (or $\cos$) with $\sinh$ (or $\sin$), it suffices to work with a positive $\gamma$ and with the upper sign in~\cref{eq:remaining_equation_general_D}.

A trivial solution to~\cref{eq:remaining_equation_general_D} exists when $\delta>0$. This has $\dot{f}=0$, for which $A$ and $\phi$ become linear functions of $\xi$. In the remaining part of this section we shall focus on the cases $\dot{f}\neq 0$.

After fixing the signs, one can write $\dot{A}$ and $\dot{\phi}$ in terms of $\psi$ as

\begin{eqaed}\label{eq:A_and_phi_general_D}
    \dot{A} & =-\frac{1}{2(D-1)}\left(\psi-\frac{1}{\psi}\right)\mcomma \\
    \dot{\phi} & = \frac{D-2}{4\sqrt{D-1}}\left(\psi+\frac{1}{\psi}\right)\mperiod
\end{eqaed}

\noindent There are three different types of solutions, depending on whether $\gamma$ is greater than, equal, or less than a critical value. In terms of $\delta$ they have the following classification:
\begin{itemize}
    \item If $\delta>0$, then
        
        \begin{eqaed}\label{eq:psi_if_delta_greater_zero}
            \psi=\frac{1}{\delta}\sqrt{\delta(\delta+1)}\tanh\left({\sqrt{\delta(\delta+1)}\xi}\right)\mcomma
        \end{eqaed}
        
        \noindent hence, from \eqref{eq:A_and_phi_general_D},
        
        \begin{eqaed}\label{eq:case_delta_greater_zero}
            A & = -\frac{1}{2(D-1)}\log\left[\left(\cosh\sqrt{\delta(\delta+1)}\xi\right)^{\frac{1}{\delta}}\left(\sinh\sqrt{\delta(\delta+1)}\xi\right)^{-\frac{1}{\delta+1}}\right]\mcomma\\
            \phi & = \varphi_0 + \frac{D-2}{4\sqrt{D-1}}\log\left[\left(\cosh\sqrt{\delta(\delta+1)}\xi\right)^{\frac{1}{\delta}}\left(\sinh\sqrt{\delta(\delta+1)}\xi\right)^{\frac{1}{\delta+1}}\right]\mperiod
        \end{eqaed}
        
    \item If $\delta =0$, then $\psi$ is linear and
        
        \begin{eqaed}\label{eq:case_delta_equal_zero}
            A & = -\frac{1}{2(D-1)}\left[\frac{1}{2}\xi^2-\log(\xi)\right]\mcomma\\
            \phi & = \varphi_0 + \frac{D-2}{4\sqrt{D-1}}\left[\frac{1}{2}\xi^2+\log(\xi)\right]\mperiod
        \end{eqaed}
        
    \item If $-\frac{1}{2}<\delta<0$, then
        
        \begin{eqaed}\label{eq:case_delta_smaller_zero}
                A & = -\frac{1}{2(D-1)}\log\left[\left(\cos\sqrt{-\delta(\delta+1)}\xi\right)^{\frac{1}{\delta}}\left(\sin\sqrt{-\delta(\delta+1)}\xi\right)^{-\frac{1}{\delta+1}}\right]\mcomma\\
                \phi & = \varphi_0 + \frac{D-2}{4\sqrt{D-1}}\log\left[\left(\cos\sqrt{-\delta(\delta+1)}\xi\right)^{\frac{1}{\delta}}\left(\sin\sqrt{-\delta(\delta+1)}\xi\right)^{\frac{1}{\delta+1}}\right]\mperiod
        \end{eqaed}
        
\end{itemize}

We recognize, for $D=10$, the orientifold solution in the critical case with $\delta=0$ and the heterotic one when $\delta=\frac{1}{3}>0$. The cosmological counterpart of what we found in this section has been analyzed in~\cite{Dudas:2010gi}, and generalizations with other scalar potentials can be found in~\cite{Fre:2013vza}.

\subsection{Critical case}\label{sec:critical_orientifold}

The dilaton potential that corresponds to the critical case has

\begin{eqaed}\label{eq:gamma_critical}
    \gamma = \gamma_c \equiv \frac{4 \sqrt{D-1}}{D-2}\mcomma 
\end{eqaed}

\noindent that is $\delta=0$, leading to~\cref{eq:case_delta_equal_zero} for metric and dilaton. After restoring all the $\alpha_{\mbox{\tiny E}}$ factors and returning to the $y$ coordinates (with an appropriate integration constant in $A$ and a redefinition of $\varphi_0$), this resembles the (critical in ten dimensions) orientifold case of~\cite{Dudas:2000ff}.
The Ricci curvature diverges at $\xi=0$ and $\xi\to\infty$, and the two timelike singularities lie at a finite distance

\begin{eqaed}\label{eq:DM_crit_length}
    L\sim e^{-\frac{1}{2}\gamma_c\varphi_0} \, ,
\end{eqaed}

\noindent leading to spontaneous compactifications for all values of $D$.

One could ask which dynamical fields survive in the reduced theory, and we shall investigate the behavior of gravity, gauge fields, and higher-form field strengths, motivated by the field content of ten-dimensional and non-critical strings. We shall postpone the more involved dilaton discussion to~\cref{sec:absence_moduli}.

As to the gravitational interaction, one must compute the Planck mass in one dimension less, that is

\begin{eqaed}\label{eq:critical_Planck_scaling}
     \frac{1}{\kappa_{D-1}^2} \propto \frac{1}{\kappa_D^2}\int^{\infty}_0 d \xi \, e^{(D-3)A+B} \propto  \frac{1}{\kappa_D^2} e^{-\frac{1}{2}\gamma_c\varphi_0}\mperiod  
\end{eqaed}

\noindent This is always finite, hence $(D-1)$-dimensional gravity is dynamical in our model. 

We now consider $p$-form field strengths, assuming Einstein-frame couplings $e^{a\phi}$. The relevant integral we must evaluate turns out to be proportional to

\begin{eqaed}\label{eq:generalized_critical_inequality_forms}
    \exp{\left(a-\frac{1}{2}\gamma_c\right)\varphi_0}\mcomma
\end{eqaed}

\noindent which converges for

\begin{eqaed}
    -4 \, \frac{D-p-1}{(D-2)\sqrt{D-1}} < a < 4 \, \frac{D-p-1}{(D-2)\sqrt{D-1}}\mperiod
\end{eqaed}

\noindent In these cases, a $(D-1)$-dimensional $p$-form field strength survives the compactification. Note the upper bound on the possible $p$-forms that are compatible with exponential dilaton couplings: $p<D-1$.

\subsection{Supercritical orientifold and heterotic}\label{sec:critical_heterotic}

We now ask the same questions for other values of $\gamma$. In particular, we shall be interested in two cases that are, in some sense, related to the orientifold and heterotic models in ten dimensions. One, that we shall call ``supercritical orientifold'', corresponds to 

\begin{eqaed}\label{eq:gerenalized_gamma_orientifold}
    \gamma_o=\frac{D+2}{D-2}\mcomma
\end{eqaed}

\noindent that we shall only consider in $D>10$, and the other, called ``supercritical heterotic'', corresponds to

\begin{eqaed}\label{eq:generalized_gamma_heterotic}
    \gamma_h=\frac{2D}{D-2}\mperiod
\end{eqaed}

\noindent We take these as natural generalizations of the ten-dimensional orientifold and heterotic cases because the corresponding ($D$-dimensional) ``string-frame'' potentials would come with $e^{-\phi}$ and no dilaton coupling, respectively. In the supercritical orientifold case, we require $D>10$ in order to fall within the supercritical solution of~\cref{eq:case_delta_greater_zero}. Note that each case consists of two possible solutions, obtained by interchanging $\sinh$ with $\cosh$. In this analysis, we shall only consider~\cref{eq:case_delta_greater_zero}, because the other solution does not have finite proper distance for any value of $\gamma$, hence no dimensional reduction occurs.

The proper length of the spatial direction singled out by the ansatz is finite for any solution from~\cref{eq:case_delta_greater_zero}, with 

\begin{eqaed}\label{eq:DM_gen_length}
    L \sim  e^{-\frac{1}{2}\gamma_{o,h}\varphi_0} \, .
\end{eqaed}

\noindent Moreover, nine-dimensional gravity is dynamical for any $D$, with the $\varphi_0$ scaling given by~\cref{eq:critical_Planck_scaling}, in which our two cases of interest simply correspond to $\gamma_c\to\gamma_{o,h}$.

$p$-form field strengths, accompanied by $e^{a\phi}$ in the action, will descend to $p$-forms in $(D-1)$ dimensions only when ($\delta_{o,h}$ is the obtained from $\gamma_{o,h}$ following~\cref{eq:nice_redefinitions})

\begin{eqaed}\label{eq:generalized_supercritical_inequality_forms}
    -4 \, \frac{D-p-1}{(D-2)\sqrt{D-1}} < a < 4 \, \frac{D-p-1+2\delta_{o,h}(\delta_{o,h}+1)(D-1)}{(D-2)\sqrt{D-1}(2\delta_{o,h}+1)}\mperiod
\end{eqaed}

\noindent Hence, $p$-form field strengths with exponential dilaton couplings are possible only for $p<(D-1)(\delta_{o,h}+1)$.

Let us now make some comments on the two special cases of~\cref{eq:gerenalized_gamma_orientifold} and~\cref{eq:generalized_gamma_heterotic}. The only differences will be in the $p$-form cases since $\gamma$ only appears there. 

Guided by gauge fields in string theory, in the supercritical orientifold case we shall consider a two-form field strength accompanied by $e^{-\phi}$ in the string frame, corresponding to $a=\frac{D-6}{D-2}$.
Then, using the appropriate value for $\delta_o$, which is valid for $D>10$,~\cref{eq:generalized_supercritical_inequality_forms} implies convergence in the range $10 < D < 26$.
Additionally, $p$-form field strengths with $a=\frac{2D-4p}{D-2}$, that would be the counterparts of R-R fields, are present in the reduced spectrum for 

\begin{eqaed}\label{eq:supercritical_orientifold_RR_range}
    \frac{3(D-2)}{8}<p<\frac{D}{2}+\frac{\sqrt{D-1}}{2}-\frac{1}{2}\mperiod
\end{eqaed}

\noindent There are always integer values for $p$ that satisfy~\cref{eq:supercritical_orientifold_RR_range}. Note that the requirement $D>10$ translates into $p>3$.

As to the supercritical heterotic case, a possible string-inspired generalization would be to consider $p$-forms with string frame $e^{-2\phi}$ couplings, that means $a=-4\frac{p-1}{D-2}$. Here, form fields in the reduced theory survive when

\begin{eqaed}
    p<\sqrt{D-1}\mperiod
\end{eqaed}

\noindent In particular, the Kalb-Ramond 3-form in ten dimensions does not survive the dimensional reduction, leaving an Einstein-Yang-Mills theory in nine dimensions.

\section[AdS compactifications with gauge fluxes]{$AdS$ compactifications with gauge fluxes}\label{sec:freund_rubin_vacua}

As we have discussed in the preceding section, integrating out the compact directions of~\cref{sec:DM_generalized}, one always obtains gravity in the remaining dimensions, while only some of the form fields survive. In this section, we consider the resulting EFT and some of its vacua, focusing on the ten-dimensional orientifold case for definiteness. Appropriate generalizations are available for the other dimensions and dilaton potentials. Before we proceed, let us remark that, although the strongly coupled edges of the Dudas-Mourad interval can be pushed to parametrically far away regions as $\varphi_0 \to -\infty$, stringy corrections could significantly affect the naive dimensional reduction\footnote{Specifically, the presence of warping and singularities needs to be treated with care. See~\cite{Lust:2022xoq} for recent efforts in this direction, albeit in a different context.}. As we will discuss in~\cref{sec:cobordisms}, there are reasons to expect that the main effect of such corrections would be to resolve the Dudas-Mourad singularity into an end-of-the-world defect. 

For $D=10$ and $\delta=0$, the ten-dimensional theory contains a Yang-Mills gauge field with Einstein-frame coupling $e^{\phi/2} \, \tr F^2$. Upon dimensional reduction in the Dudas-Mourad background, the gauge coupling is finite~\cite{Dudas:2000ff}, and fixing the Planck length $\ell_\text{Pl} \propto (\alpha')^\frac{1}{2} e^{\frac{\gamma_c}{14} \phi_0}$ one is left with gravity and gauge fields in the nine-dimensional EFT described by the action

\begin{eqaed}\label{eq:reduced_action_DM}
    S\sim \frac{1}{\ell_\text{Pl}^{7}} \int d^9 x \,  \sqrt{g}\left[R-\frac{\ell_\text{Pl}^2}{2}e^{\frac{2}{7} \varphi_0}\tr F^2\right]\mcomma
\end{eqaed}

\noindent namely nine-dimensional Einstein-Yang-Mills theory, where we have rescaled $F$ with an $\mathcal{O}(1)$ factor from the dimensional reduction.
The simplest types of vacua one can extract from this action consist of gauge fields with U(1) vacuum values in the Cartan subalgebra and Freund-Rubin spacetimes that are products of a two-dimensional manifold times a seven-dimensional one, as in~\cite{Raucci:2022bjw}.
The Yang-Mills equations then require electric or magnetic field strengths, proportional to the 2-volume form. Because of this structure, we will treat the gauge field as if it were abelian, thereby suppressing Lie-algebra indices.

The first solution is an $AdS_2 \times S^7$ compactification with the round metric on the $S^7$, although any other compact Einstein manifold with appropriate constant curvature yields a similar solution. The vacuum is completely characterized by

\begin{eqaed}
    F_{\mu\nu}  = f \, \epsilon_{\mu\nu}\mcomma\qquad
    R_{\mu\nu}  = -\frac{3}{7}f^2 \ell_\text{Pl}^2 e^{\frac{2}{7}\varphi_0}g_{\mu\nu}\mcomma\qquad
    R_{mn}  = \frac{1}{14}f^2 \ell_\text{Pl}^2 e^{\frac{2}{7}\varphi_0}g_{mn}\mperiod
\end{eqaed}

\noindent As usual in Freund-Rubin vacua, the field equations require that $f$ be constant, and its value can be re-expressed in terms of the electric flux number $N_e \propto \int_{S^7} \star e^{\frac{2}{7} \phi_0} F_2$. Up to irrelevant $\mathcal{O}(1)$ numerical factors, the two $AdS_2$ and $S^7$ curvature radii scale according to

\begin{eqaed}\label{eq:gauge_vacua_electric_radii}
    R_{S^7}\sim l_{AdS_2}\sim \ell_\text{Pl} \, e^{-\frac{1}{42}\varphi_0} \, N_e^{\frac{1}{6}}\mperiod
\end{eqaed}

\noindent Analogously, the second possibility is an $AdS_7 \times S^2$ compactification with a magnetic U(1) flux $N_m \propto \int_{S^2} F_2$ threading the $S^2$. This leads to a Freund-Rubin vacuum whose curvature radii have a linear dependence on the magnetic flux number,

\begin{eqaed}\label{eq:gauge_vacua_magnetic_radii}
    R_{S^2}\sim l_{AdS_7}\sim \ell_\text{Pl} \, e^{\frac{1}{7}\varphi_0}N_m\mperiod
\end{eqaed}

All in all, these solutions are not qualitatively different from other non-supersymmetric Freund-Rubin vacua, such as the ten-dimensional studied in~\cite{Gubser:2001zr, Mourad:2016xbk, Basile:2018irz, Antonelli:2019nar}. Among the shared features, the field equations force spacetime to be $AdS$, and they forbid scale separation according to~\cref{eq:gauge_vacua_electric_radii} and its magnetic counterpart~\cref{eq:gauge_vacua_magnetic_radii}. These features are expected on general grounds, both from several examples and from swampland considerations. Another similarity that these simpler nine-dimensional vacua share with their ten-dimensional counterparts is that they are parametrically weakly curved for large fluxes. Actually, in this case, the weakly coupled parameter space is controlled both by $\varphi_0$ and the flux $N$, since we expect these solutions to be reliable whenever

\begin{eqaed}\label{eq:small_curvature_limits}
    e^{\varphi_0} \ll 1 \, , \qquad e^{\mp \frac{1}{7}\varphi_0} \, N_{e,m} \gg 1
\end{eqaed}

\noindent including both types of fluxes. Similar results can be derived for the heterotic model or more general dilaton potentials. As we will discuss in~\cref{sec:moduli_kk}, this is consistent with the expectation that $\varphi_0$ is determined by the number of 8-branes sourcing the Dudas-Mourad geometry. Indeed, $\varphi_0$ cannot be interpreted as a modulus arising from the dimensional reduction over the interval.

As a final comment, let us observe that the nine-dimensional theory we have discussed clearly allows scale-separated Ricci flat compactifications. However, they bring along moduli with no obvious novel stabilization mechanism. More generally, the results we have presented in this section illustrate a recurring theme in string-scale supersymmetry breaking: within the regime of validity of EFTs, the issues of scale separation and the (sign of the) cosmological constant remain even after supersymmetry is broken. In fact, one could think of this as an indication that these models are compatible with swampland conditions, and indeed the very starting point of our constructions avoids the arguments of~\cite{Cribiori:2021gbf, Castellano:2021yye, DallAgata:2021nnr}. As a further check of consistency with swampland conditions, in~\cref{sec:decomp_limit} we show that KK modes arising from the Dudas-Mourad interval behave in a manner consistent with the distance conjecture with a specific decay rate constant.

\subsection{Indications of (in)stability}\label{sec:perturbations_FR_vacua}

We now begin the investigation of perturbative stability for the two Freund-Rubin vacua, focusing on a subset of all possible perturbations. 

For the time being, let us consider metric perturbations $h_{MN}$ and gauge perturbations $a_M$ only along the background U(1). This is consistent because, at the linearized level, non-abelian modes do not mix with the other types of perturbations. The linearized equations become

\begin{eqaed}\label{eq:linearized_eom_FR_vacua}
    \Box h_{MN} & + \nabla_M\nabla_N h -2\nabla_{(M}(\nabla\cdot h)_{N)} -2 {R^{B}}_{(M}{h_{N)B}}+2{R^B}_{MAN} {h_B}^A +\\
        & + \ell_\text{Pl}^2 e^{\frac{2}{7}\varphi_0}\left[2{F_M}^A \nabla_{[N}a_{A]}+2{F_N}^A \nabla_{[M}a_{A]}-F_{MA}F_{NB}h^{AB}\right]+\\
            & -\frac{\ell_\text{Pl}^2}{14}e^{\frac{2}{7}\varphi_0}g_{MN}\left[4F_{AB}\nabla^A a^B -2F_{AB}{F_C}^B h^{AC}\right]+\\
                &-\frac{\ell_\text{Pl}^2}{14}e^{\frac{2}{7}\varphi_0}F^2 h_{MN}=0 \mcomma\\
    \Box a^N & -\nabla_M \nabla^N a^M -F^{AN} (\nabla\cdot h)_A-F^{MA}\nabla_M {h_A}^N+\frac{1}{2}F^{MN}\nabla_M h=0\mperiod
\end{eqaed}

Let us simplify our notation, letting

\begin{eqaed}\label{eq:linearized_units}
    \tau=\frac{\ell_\text{Pl}^2}{2}e^{\frac{2}{7} \varphi_0}f^2
\end{eqaed}

\noindent and

\begin{eqaed}\label{eq:linearized_L}
    L=\frac{l(l+6)}{42} \qquad \mbox{or} \qquad \frac{6l(l+1)}{7}\mcomma
\end{eqaed}

\noindent where we employ the former notation for the $AdS_2$ solution and the latter for the $AdS_7$ one.
We classify perturbations in terms of their behavior under the isometries of the background, and we implicitly expand the fields in terms of spherical harmonics on the spheres. We also denote the $9D$, $AdS$ and internal Laplacian operators as $\Box$, $\Box_{AdS}$ and $\nabla^2$, so that $L\tau$ represents the eigenvalues of the internal Laplacian on spherical harmonics.
In both cases, tensor modes in $AdS$ are described by

\begin{eqaed}\label{eq:linearized_tensor_modes}
    \Box h_{\mu\nu}=-\frac{2}{l_{AdS}^2}h_{\mu\nu}\mcomma
\end{eqaed}

\noindent representing a massless graviton and its KK tower, therefore bringing no instability (for $AdS_2$ these are pure gauge).
Similarly, $h_{ij}$ fluctuations, that are $AdS$ scalars and tensors with respect to the internal rotation group, are stable and satisfy

\begin{eqaed}\label{eq:linearized_hii}
    \Box_{AdS} h_{ij}=L\tau h_{ij}\mperiod
\end{eqaed}

Let us consider the remaining modes in the $AdS_2\times S^7$ case. Internal vectors arise from

\begin{eqaed}\label{eq:AdS2_vectors_def}
    h_{\mu i}=\epsilon_{\mu\nu}\nabla^\nu V_i \qquad \mbox{and} \qquad a_i\mcomma
\end{eqaed}

\noindent and mix according to

\begin{eqaed}\label{eq:AdS2_vectors_eom}
    \Box_{AdS} \begin{pmatrix} V_i \\ \frac{1}{f}a_i \end{pmatrix} = \begin{pmatrix}L-\frac{1}{6} & 2\\ L-\frac{1}{6} & L+\frac{89}{42}\end{pmatrix}\tau \begin{pmatrix} V_i \\ \frac{1}{f}a_i \end{pmatrix}\mperiod
\end{eqaed}

\noindent The resulting mass matrix has a vanishing eigenvalue for $l=1$ and positive eigenvalues for all other cases, and therefore this sector is stable.

Singlet scalar modes correspond to fluctuations

\begin{eqaed}\label{eq:AdS2_singlet_scalars}
    h_{\mu\nu}=Ag_{\mu\nu}\mcomma \qquad h_{ij}=C g_{ij}\mcomma \qquad h_{\mu i} =\frac{1}{\tau}\nabla_\mu\nabla_i D\mcomma \qquad a_\mu=\epsilon_{\mu\nu}\nabla^\nu a\mcomma 
\end{eqaed}

\noindent which parameterize our fields up to gauge transformations with independent parameters on the two manifolds. For $l=0$ there is no dependence on the internal sphere, and it is possible to express these modes in terms of $A$ and $a$, with a mixing matrix whose eigenvalues lie above the B-F bound. A similar behavior emerges for $l>0$, where two algebraic relations among the linearized equations bring the independent fluctuations to the form

\begin{eqaed}\label{eq:AdS2_singlet_scalars_eom}
    \Box_{AdS} \begin{pmatrix} A \\  D \end{pmatrix}=\begin{pmatrix} \frac{19}{7}L+\frac{12}{7} & \frac{24}{49}L(1-6L) \\ 1 & -\frac{5}{7}L \end{pmatrix}\tau \begin{pmatrix}A \\  D \end{pmatrix}\mperiod
\end{eqaed}

\noindent One branch of the two eigenvalues can become negative, being $\frac{l(l-6)}{42}\tau$, but, after adding the appropriate B-F bound, all modes are stable.\footnote{Fluctuations with $l=3$ are marginally stable, since the eigenvalue exactly matches the B-F bound.}

We now turn to the other $AdS_7\times S^2$ type of vacua. Vector modes from 

\begin{eqaed}\label{eq:AdS7_vectors_def}
    h_{\mu i}=\epsilon_{ij}\nabla^j V_\mu \qquad \mbox{and} \qquad a_\mu
\end{eqaed}

\noindent mix according to

\begin{eqaed}\label{eq:AdS7_vectors_eom}
    \Box_{AdS} \begin{pmatrix} V_\mu \\ \frac{1}{f}a_\mu \end{pmatrix} = \begin{pmatrix}L + \frac{1}{7} & 2 \\ L & L-\frac{1}{7}\end{pmatrix}\tau \begin{pmatrix} V_\mu \\ \frac{1}{f}a_\mu \end{pmatrix}\mperiod
\end{eqaed}

\noindent For $l=0$ only the $a_\mu$ modes are present, subject to the massless $AdS$ Maxwell equations. For $l=1$ there is a triplet of massless vectors representing the isometries of $S^2$, while all other modes have positive squared masses.

Singlet scalar perturbations can be parameterized as in~\cref{eq:AdS2_singlet_scalars}, with the only difference being $a_i=\epsilon_{ij}\nabla^j a$ instead of $a_\mu$. The $l=0$ subsector is again simpler, because $a$ and $D$ are absent, and it is possible to write these modes in terms of a single equation for $C$, with a positive definite spectrum.
In the generic $l\neq0$ case, one can use the two resulting algebraic equations in order to reduce scalar perturbations to

\begin{eqaed}\label{eq:AdS7_singlet_scalars_eom}
    \Box_{AdS} \begin{pmatrix} \frac{\tau}{f}a \\ C \end{pmatrix}=\begin{pmatrix} L & -1 \\ -\frac{24}{7}L & L+\frac{12}{7} \end{pmatrix}\tau \begin{pmatrix}\frac{\tau}{f}a \\ C \end{pmatrix}\mperiod
\end{eqaed}

\noindent The eigenvalues of this mass matrix lie above the B-F bound, therefore this sector is stable.

In the above analysis, we found no explicit example of unstable modes, but we expect that both types of vacua be unstable, because non-abelian gauge perturbations can mix with the background U(1). In fact, it is a known result in flat space that constant electric or magnetic fields result in unstable modes when the fluxes are large enough~\cite{Chang:1979tg, Sikivie:1979bq}. In the regime of validity of our vacua, encoded in~\cref{eq:small_curvature_limits}, curvatures are small and fluxes are large, and therefore we expect the same mechanism to play a role. A detailed analysis of this type of gauge instability in $AdS\times S$ backgrounds will be the subject of future investigations. 

All in all, our analysis in this section delimits the presence of (perturbative) instabilities to the non-abelian gauge sector, since no unstable modes emerge in the gravitational and abelian sector. Even if unstable modes were present, as expected on general grounds, it is conceivable that replacing the internal sphere with a suitable compact Einstein manifold or orbifold\footnote{This has been discussed in~\cite{Basile:2018irz} for ten-dimensional heterotic vacua with R-R fluxes.} could rid these vacua of instabilities. The ultimate fate of these settings would then be determined by non-perturbative instabilities, which in this case could comprise bubbles of nothing with gauge flux attached, or transitions mediated by Yang-Mills instantons. It would be interesting to compare these effects to the analyses of~\cite{Antonelli:2019nar,Guarino:2020flh, Bomans:2021ara, Guarino:2021hrc, Giambrone:2021wsm, Guarino:2022tlw, Dibitetto:2022rzy}.

\section{Moduli and Kaluza-Klein masses}\label{sec:moduli_kk}

In this section, we argue that the free parameter $\varphi_0$ in the Dudas-Mourad solutions cannot be interpreted as a modulus. To this end, we provide two independent checks. The most substantial includes the analysis of the KK spectrum of the Dudas-Mourad geometry, which allows us to conclude that moduli and perturbative instabilities are absent in all dimensions $D$. At the level of field fluctuations around the solution, the ten-dimensional solutions of~\cite{Dudas:2000ff} are already known to be perturbatively stable~\cite{Basile:2018irz}. The only potentially offending perturbations are scalars, which can be reduced to a single eigenvalue problem. While the full spectrum of KK modes on the internal interval was not worked out in~\cite{Basile:2018irz}, the positivity of the squared masses follows from a simpler argument, which we now revisit and extend.

\subsection{Perturbative stability}\label{sec:DM_pert_stability}

Here we extend these considerations to the general settings presented in~\cref{sec:DM_generalized}. To this end, following the approach in~\cite{Basile:2018irz}, we choose conformally flat coordinates $x^M = (x^\mu, z)$ such that the solution takes the form

\begin{eqaed}\label{eq:9d_vacuum}
    & ds^2 = e^{2\Omega(z)} \, \eta_{MN} \, dx^M dx^N \, , \nonumber \\
    & \phi = \phi_0(z) \, , \nonumber \\
\end{eqaed}

\noindent and once again only the scalar perturbations

\begin{eqaed}\label{eq:9d_scalar_pert}
    & ds^2 = e^{2\Omega(z)} \, (\eta_{MN} + h_{MN}(x,z)) \, dx^M dx^N \, , \nonumber\\
    & h_{\mu \nu} = A \, \eta_{\mu \nu} \, , \nonumber \\
    & h_{\mu z} = \partial_\mu \, \widehat{D} \, , \nonumber \\
    & h_{z z} = C \, , \nonumber\\
    & \phi = \phi_0(z) + \varphi(x,z)
\end{eqaed}

\noindent can lead to perturbative instabilities. To lighten the notation, we shall write $V_0 \equiv V(\phi_0)$ and denote derivatives with respect to $z$ with primes, except for the potential $V$ where primes denote derivatives with respect to $\phi$. The resulting linearized field equations read 

\begin{eqaed}\label{eq:9d_linearized}
   & -\frac{1}{2} \, \left(\Box \, h_{MN} - \partial_{(M} \partial \cdot h_{N)} + \partial_M \partial_N h_A^A \right) + \frac{D-2}{2} \, \Omega' \left( \partial_{(M} h^z_{N)} - \partial^z h_{MN} \right) \nonumber\\
   & - \left(\Omega'' + (D-2){\Omega'}^2 \right) \, h_{MN} + \left( \left(\Omega'' + (D-2){\Omega'}^2 \right) \, h^{zz} + \Omega' \, \partial \cdot h^z - \frac{\Omega'}{2} \, \partial^z h_A^A  \right) \, \eta_{MN} \nonumber\\
   & = \frac{4}{D-2}\, \partial_{(M} \phi_0 \partial_{N)} \varphi + \frac{e^{2\Omega}}{D-2} \left(V_0 \, h_{MN} + V_0' \, \varphi \, \eta_{MN} \right) \, , \nonumber\\
   & \Box \, \varphi - h^{zz} \, \phi_0'' - \phi_0' \, \left( \partial \cdot h^z - \frac{1}{2} \, \partial^z h_A^A \right) - (D-2) {\Omega'}^2 \, h^{zz} = \frac{D-2}{8} \, e^{2\Omega} \, V_0'' \, \varphi \, ,
\end{eqaed}

\noindent and for the scalar perturbations in~\cref{eq:9d_scalar_pert} they simplify to

\begin{eqaed}\label{eq:9d_scalar}
   & - \frac{1}{2} \, \Box \, A - \frac{2D-3}{2} \, \Omega' \, A' + \left(\Omega'' + (D-2){\Omega'}^2 \right) \, C + \frac{\Omega'}{2} \, C' + \Omega' \, \partial^2 \, \widehat{D} = \frac{e^{2\Omega}}{D-2} V_0' \, \varphi \, , \nonumber\\
   & (D-3) \, A + C - 2(D-2) \Omega' \, \widehat{D} - 2 \widehat{D}' = 0 \, , \nonumber\\
   & - \frac{1}{2} \partial^2 \, C + \partial^2 \, \widehat{D}' - \frac{D-1}{2} \, A'' + \frac{D-1}{2} \, \Omega' \, C' + \Omega' \partial^2 \, \widehat{D} \nonumber\\
   & - \frac{D-1}{2} \, \Omega' \, A' = \frac{8}{D-2} \, \phi_0' \, \varphi' + \frac{e^{2\Omega}}{D-2} \left( V_0 \, C + V_0' \, \varphi \right) \, , \nonumber\\
   & - \frac{D-2}{2} \, A' + \frac{D-2}{2} \, \Omega' \, C = \frac{4}{D-2} \, \phi_0' \, \varphi \, , \nonumber\\
   & \Box \, \varphi - \phi_0'' \, C - \phi_0' \left( \frac{1}{2} \, C' + \partial^2 \, \widehat{D} - \frac{D-1}{2} \, A' \right) - (D-2) {\Omega'}^2 \, C = \frac{D-2}{8} \, e^{2\Omega} \, V_0'' \, \varphi \, ,
\end{eqaed}

\noindent where $\partial^2 \equiv \eta_{\mu \nu} \, \partial_\mu \, \partial_\nu$ and we used the vacuum equations to simplify some terms.

One can conveniently use the diffeomorphism invariance along $z$ to gauge away $\widehat{D}$. Hence, one finds

\begin{eqaed}\label{eq:A_C_constraint}
    C = -(D-3) \, A
\end{eqaed}

\noindent which, along with the other constraints, allows to isolate $A$ as the only scalar degree of freedom of the system. Indeed, isolating

\begin{eqaed}\label{eq:dilaton_isolation}
    \varphi = - \frac{(D-2)^2}{8 \, \phi_0'} \left( A' + (D-3) \Omega' \, A \right)
\end{eqaed}

\noindent one finds

\begin{eqaed}\label{eq:A_final_eq}
    \Box \, A + (D-2) \left( 3 \, \Omega'  -\frac{e^{2\Omega}V_0'}{4\, \phi_0'} \right) A' - 2 (D-3) \, e^{2\Omega} \left(\frac{V_0}{D-2} + (D-2) \frac{\Omega' V_0'}{8 \, \phi_0'} \right) A = 0 \, .
\end{eqaed}

\noindent This is the ``master equation'' describing linearized scalar perturbations. One can perform a Fourier transform to expose the $D-1$ translational modes, replacing them with momenta $p^\mu$ with $p^2 = - \, m^2$. Then, $\Box A \to A'' + m^2 \, A$. One can then recast the resulting equation in a Schr\"{o}dinger-like form via the substitution

\begin{eqaed}
    A = \Psi \, \exp \left[ -\frac{D-2}{2} \int_{z_0}^z \left( 3 \, \Omega'  -\frac{e^{2\Omega}V_0'}{4\, \phi_0'} \right) ds \right] \, .
\end{eqaed}

\noindent Analogously to the special case in~\cite{Basile:2018irz}, the equation takes the guise of a Schr\"{o}dinger eigenvalue problem for the squared masses $m^2 = - \eta_{\mu \nu} p^\mu \, p^\nu$,

\begin{eqaed}\label{eq:eig_prob}
    \mathcal{H} \, \Psi = m^2 \Psi \, ,
\end{eqaed}

\noindent where the ``Hamiltonian'' $\mathcal{H}$ can be written in the form~\cite{Basile:2018irz}

\begin{eqaed}\label{eq:hamiltonian}
    \mathcal{H} \equiv b + \mathcal{A}^\dagger \, \mathcal{A} \, ,
\end{eqaed}

\noindent where the annihilation-like operator

\begin{eqaed}\label{eq:ladder_like}
    \mathcal{A} \equiv -\frac{d}{dz} + \frac{D-2}{2} \left( 3 \, \Omega'  -\frac{e^{2\Omega}V_0'}{4\, \phi_0'} \right)
\end{eqaed}

\noindent has the adjoint

\begin{eqaed}\label{eq:adjoint_ladder_like}
    \mathcal{A}^\dagger \equiv \frac{d}{dz} + \frac{D-2}{2} \left( 3 \, \Omega'  -\frac{e^{2\Omega}V_0'}{4\, \phi_0'} \right)
\end{eqaed}

\noindent with Dirichlet or Neumann boundary conditions\footnote{We recall that the extent of the $z$ direction is finite.}. Finally, the function $b$ is

\begin{eqaed}\label{eq:b_function}
    b \equiv 2 (D-3) \, e^{2\Omega} \left(\frac{V_0}{D-2} + (D-2) \frac{\Omega' V_0'}{8 \, \phi_0'} \right) \, .
\end{eqaed}

\noindent Hence, the spectrum contains non-negative squared masses for $b \geq 0$, or equivalently

\begin{eqaed}\label{eq:stability_cond_9d}
    \frac{\Omega'(z)}{\phi_0'(z)} = \frac{dA/dy}{d\phi_0/dy} \geq - \, \frac{8}{\gamma (D-2)^2} \, ,
\end{eqaed}

\noindent where we used $V_0' = \gamma \, V_0 > 0$. This condition was verified for both the orientifold and heterotic vacua in ten dimensions~\cite{Basile:2018irz}. Substituting~\eqref{eq:A_and_phi_general_D} in~\eqref{eq:stability_cond_9d}, one arrives at

\begin{eqaed}\label{eq:final_stability_cond}
    \frac{1}{\psi}\geq \delta \left(\psi-\frac{1}{\psi}\right)\mcomma
\end{eqaed}

\noindent which, using the differential equation for $\psi$, becomes $\dot{\psi}\geq0$. This is always true for both the critical case of~\cref{sec:critical_orientifold}, in which $\dot{\psi}=1$, and the supercritical cases of~\cref{sec:critical_heterotic}, as can be verified from~\cref{eq:psi_if_delta_greater_zero}.

The above argument suffices to conclude perturbative stability, at least at the two-derivative EFT level. Even if this feature survives higher-derivative corrections, the vacua are expected to be metastable at best~\cite{Ooguri:2016pdq, Freivogel:2016qwc, Antonelli:2019nar, Dibitetto:2020csn, GarciaEtxebarria:2020xsr}. Bubbles of nothing appear to be the best candidates for a decay channel, but they have not been found yet in this case.

\subsection{Decompactification limit}\label{sec:decomp_limit}

At any rate, it is important to push this analysis slightly deeper: in addition to stability properties, the masses of KK states encode other interesting physics. In particular, they constitute an infinite tower of states, which becomes light in the decompactification limit. According to the distance conjecture~\cite{Ooguri:2006in}, this decay should be exponential in the field excursion $\Delta \phi$ in scalar field space, in the sense that

\begin{eqaed}\label{eq:kk_sdc}
    \frac{m_{\text{KK}}(\phi)}{m_{\text{KK}}(\phi_0)} \overset{\Delta \phi \to \infty}{\sim} e^{- \mathcal{O}(1) \Delta \phi} \, .
\end{eqaed}

\noindent Let us now see how this occurs in the present setting. To begin with, recall that the Dudas-Mourad geometries are characterized by a single free parameter $\varphi_0$, which arises from the $D$-dimensional profile of the dilaton. Because of our gauge choice $B = - \frac{\gamma}{2} \phi$, the solution written in terms of the $y$ coordinate depends on $\varphi_0$ only additively in $B(y)$ and $\phi(y)$. Therefore, in the conformally flat coordinates of~\cref{eq:9d_vacuum} $dz = e^{B-A} \, dy$ scales as $e^{-\frac{\gamma}{2} \varphi_0}$, while $\partial_z$ scales as its reciprocal, As a result, upon substituting $\Box A \to A'' + m^2 \, A$ in~\cref{eq:A_final_eq}, all terms scale as $e^{\gamma \varphi_0}$ except $m^2 \, A$. Hence, one can rewrite~\cref{eq:A_final_eq} in terms of the single free eigenvalue $m^2 e^{-\gamma \varphi_0}$, which generically assumes $\mathcal{O}(1)$ positive values. Thus, varying $\varphi_0 \to \varphi$, the same eigenvalue varies according to

\begin{eqaed}\label{eq:kk_sdc_proof}
    \frac{m^2(\varphi)}{m^2(\varphi_0)} = e^{\gamma \Delta \varphi} \, ,
\end{eqaed}

\noindent which corresponds to an exponential decay for $\varphi \to -\infty$. This is indeed the decompactification limit~\cite{Basile:2020xwi, Basile:2021vxh}, as evident from~\cref{eq:DM_gen_length}. In particular, combining~\cref{eq:DM_gen_length} with~\cref{eq:kk_sdc_proof} shows that the Einstein-frame proper length $L$ of the internal dimension scales as $\frac{1}{m}$ as expected. Here, the $\mathcal{O}(1)$ constant of~\cref{eq:kk_sdc} is fixed to $\frac{\gamma}{2}$.

\subsection{Absence of moduli}\label{sec:absence_moduli}

According to the preceding discussion, one could be tempted to conclude that there is an associated modulus in the $(D-1)$-dimensional EFT, whose VEV gives back $\varphi_0$. This is actually not the case, since there is no normalizable zero-mode solution of~\cref{eq:A_final_eq}. The potential closely resembles an infinite well of length $L$, the proper length of the internal interval. Thus, one can expect at least the first few eigenvalues to be approximately $m_n \approx n \, \frac{\pi}{L}$, with $n>0$. Indeed, a numerical analysis of~\cref{eq:A_final_eq} and~\cref{eq:eig_prob} based on the shooting method in the ten-dimensional orientifold models reveals that the first KK masses $m_n$ occur at

\begin{eqaed}\label{eq:kk_gap_orientifold}
    m_1 \approx 0.97 \, \frac{\pi}{L} \, , \qquad m_2 \approx 2.13 \, \frac{\pi}{L} \, .
\end{eqaed}

\noindent Accordingly, the even (odd) profile $\Psi_1$ ($\Psi_2$) resembles the corresponding sinusoidal solution of a quantum particle in a box. Of course, we do not expect this intuition to persist for very massive KK states since the corresponding eigenfunctions would probe the deviation of the potential from the infinite well more accurately. We have not been able to extract eigenvalues with similar precision in the heterotic model due to numerical instabilites, although the qualitative behavior is similar.

For completeness, we mention that additional indications that $\varphi_0$ should not be taken as a modulus come from the equations of motion. In fact, promoting $\varphi_0$ to a spacetime-dependent field $\varphi(x)$ has the consequence of allowing a non-vanishing $\mu y$ gravitational equation. In any dimension, since $A'$ and $\phi'$ have different functional forms in $y$, this implies that

\begin{eqaed}
    \d_\mu\varphi=0\mperiod
\end{eqaed}

\noindent Therefore, only a constant value for $\varphi_0$ is allowed, ruling out any interpretation as a dynamical modulus.

Despite the absence of moduli, the exponential vanishing of KK masses remains physically significant: it is natural to expect $\varphi_0$ to play a role akin to parameters such as the number of branes in similar configurations. Some evidence in favor of this was found in~\cite{Antonelli:2019nar}, where the Dudas-Mourad solutions were connected to a more general family of $p$-brane solutions, and the free parameter in the extremal case is indeed connected to the number of brane sources via a Dirac quantization condition. Accordingly, the scaling of~\cref{eq:kk_sdc_proof} would resonate with generalized versions of the distance conjecture~\cite{Lust:2019zwm, Baume:2020dqd, Perlmutter:2020buo}, perhaps one that can encompass discrete parameters, such as the one put forth in~\cite{Basile:2022zee}.

\subsection{On cobordisms to nothing}\label{sec:cobordisms}

Let us conclude this analysis with a few speculative remarks. The solutions that we have discussed in this paper feature yet another example of what has been dubbed ``(local) dynamical cobordism''~\cite{Buratti:2021fiv, Buratti:2021yia, Angius:2022aeq}, since spacetime ends in a singularity at finite distance. This type of behavior has been connected to the distance conjecture and the cobordism conjecture~\cite{McNamara:2019rup}, which has recently received much attention~\cite{Blumenhagen:2021nmi, Andriot:2022mri, Blumenhagen:2022mqw} and connects topology change to (Dai-Freed) anomalies~\cite{Garcia-Etxebarria:2018ajm, Debray:2021vob}.

It is conceivable that the infinite-distance scaling of the KK masses in~\cref{eq:kk_sdc_proof} could be connected to the existence of an end-of-the-world (ETW) defect, as advocated in~\cite{Buratti:2021fiv, Buratti:2021yia, Angius:2022aeq}. At the $D$-dimensional level, one can clearly see the diverging field excursion, and the appearance of the ETW defect seems to be universal. To wit, replacing the putative 8-brane source of the Dudas-Mourad geometry with general (charged or uncharged) $p$-brane sources, the codimension-one ETW defect still appears, albeit wrapped around the dimensions transverse to the $p$-branes~\cite{Antonelli:2019nar}. This supports the idea that the ETW defect is an actual physical object, and indeed in the T-dual picture of~\cite{Blumenhagen:2000dc}, where one has more transverse dimensions, a solution for the isolated ETW 7-brane was recently found~\cite{Blumenhagen:2022mqw}. While it is unclear whether a similar approach can be understood directly from our duality frame, the classification of bordism groups stemming from~\cite{Debray:2021vob} can probe the existence of suitable 8-brane and 7-brane defects connected by T-duality\footnote{We thank Arun Debray for pointing this out to us.}.

From the $(D-1)$-dimensional perspective, the end of spacetime should manifest itself as a domain wall to nothing. In order for this to be possible in all configurations of the theory, the appropriate bordism group ought to vanish. It is unclear what this appropriate structure could be\footnote{See~\cite{Andriot:2022mri} for a possible connection to the Whitehead tower.}, but one can expect these arguments to remain valid in the absence of supersymmetry, since they do not rely on it. This is particularly advantageous in the settings that we have discussed in this paper, since it is not clear whether a fully stable semiclassical configuration even exists\footnote{There are however some hints that these theories could survive in a stringy regime living at infinite distance~\cite{Basile:2022zee}. Hence, there could be a weakly coupled dual frame describing the physics.} -- a topological argument to establish or rule out the consistency of these models could overcome this difficulty. 

\section{Conclusions}\label{sec:conclusions}

    In this paper, we revisited and clarified several aspects of the Dudas-Mourad geometry. In particular, we focused on dimensional reduction, which turns out to yield Einstein-Yang-Mills theory at low energies. Furthermore, we showed that this setup contains no moduli. However, the full geometry is \emph{a priori} unreliable globally, due to the singularities at the endpoints of the interval. We discussed a possible connection between this issue, the triviality of cobordism classes and dynamical tadpoles.
    
    The main upshot of our analysis is that the spontaneous compactification driven by dynamical gravitational tadpoles leads to simpler EFTs, devoid of higher-form fields and scalar moduli (and of course supersymmetry) at low energies. However, this type of dimensional reduction, if at all reliable due to the issues discussed in~\cref{sec:cobordisms}, only works when reducing from 10d to 9d. Additional compactifications are fraught with the standard trade-off between scale separation and moduli stabilization, as we discussed in~\cref{sec:freund_rubin_vacua}. In this context, the spontaneous emergence of a privileged compact dimension is somewhat reminiscent of the recent ``dark dimension'' scenario proposed in~\cite{Montero:2022prj} (see also~\cite{Anchordoqui:2022txe, Blumenhagen:2022zzw}).
    
    A workaround could be studying non-critical string theories applying the methods of~\cref{sec:DM_generalized}. It is nonetheless conceivable that string corrections would be relevant in any scenario of this type. A cobordism-based kinematical approach could be instructive in this respect: by identifying intrinsic properties of the putative end-of-the-world defects responsible for the classical singularities, one could devise methods to extract more reliable lessons from string-scale supersymmetry breaking. For instance, the approach of~\cite{Debray:2021vob} suggests examining the geometry from different (T-)dual frames, a direction initially explored in~\cite{Blumenhagen:2000dc, Dudas:2002dg} and recently revisited in~\cite{Blumenhagen:2022mqw}. Analyzing the scaling properties of the singularities~\cite{Antonelli:2019nar, Buratti:2021fiv, Buratti:2021yia, Angius:2022aeq} in each frame could provide further insight into their stringy nature. This would potentially guide investigations beyond the low-energy EFT, or show an inconsistency of these backgrounds in string theory. If such an inconsistency were to be found, braneworlds could provide an alternative to scale-separated compactifications. In supersymmetric settings one can show how effective supergravity dynamics arises on the worldvolume~\cite{Stelle:2020mmg, Erickson:2021psj, Leung:2022nhy}. With (string-scale) supersymmetry breaking, however, one naturally finds nucleating branes in metastable AdS vacua~\cite{Antonelli:2019nar}, yielding dS braneworld cosmologies~\cite{Basile:2020mpt}. These provide the first known string construction of the dS bubble scenario studied by~\cite{Banerjee:2019fzz, Banerjee:2020wix, Banerjee:2020wov, Banerjee:2021qei, Banerjee:2021yrb, Danielsson:2021tyb, Danielsson:2022fhd} from a bottom-up perspective.

\section*{Acknowledgements}

    The work of I.B.\ was supported by the Fonds de la Recherche Scientifique - FNRS under Grants No.\ F.4503.20 (``HighSpinSymm'') and T.0022.19 (``Fundamental issues in extended gravitational theories''). The work of S.R. was supported in part by Scuola Normale Superiore, by INFN (IS GSS-Pi) and by the MIUR-PRIN contract 2017CC72MK\_003.
    I.B. would like to thank Ralph Blumenhagen, Andriana Makridou, Niccolò Cribiori, Christian Kneissl, Matilda Delgado, José Calderón-Infante, Miguel Montero and Arun Debray for discussions during the conference ``Strings and Geometry 2022'' and the workshop ``Geometric Aspects of the Swampland''. S.R. would like to thank Augusto Sagnotti for highlighting relevant references.

% ---------------------------------------------------------------------------------------------------------------------------------
%		Bibliography
% ---------------------------------------------------------------------------------------------------------------------------------

\printbibliography

@article{Blum:1997cs,
    author = "Blum, Julie D. and Dienes, Keith R.",
    title = "{Duality without supersymmetry: The Case of the SO(16) x SO(16) string}",
    eprint = "hep-th/9707148",
    archivePrefix = "arXiv",
    reportNumber = "IASSNS-HEP-97-67",
    doi = "10.1016/S0370-2693(97)01172-6",
    journal = "Phys. Lett. B",
    volume = "414",
    pages = "260--268",
    year = "1997"
}

@article{Blum:1997gw,
    author = "Blum, Julie D. and Dienes, Keith R.",
    title = "{Strong / weak coupling duality relations for nonsupersymmetric string theories}",
    eprint = "hep-th/9707160",
    archivePrefix = "arXiv",
    reportNumber = "IASSNS-HEP-97-80",
    doi = "10.1016/S0550-3213(97)00803-1",
    journal = "Nucl. Phys. B",
    volume = "516",
    pages = "83--159",
    year = "1998"
}

@article{Dienes:1998wu,
    author = "Dienes, Keith R.",
    editor = "Derendinger, J. P. and Lucchesi, C.",
    title = "{Duality without supersymmetry}",
    reportNumber = "CERN-TH-98-06, CERN-TH-98-006",
    journal = "Fortsch. Phys.",
    volume = "47",
    pages = "141--149",
    year = "1999"
}

@article{Angelantonj:2007ts,
    author = "Angelantonj, Carlo and Dudas, Emilian",
    title = "{Metastable string vacua}",
    eprint = "0704.2553",
    archivePrefix = "arXiv",
    primaryClass = "hep-th",
    reportNumber = "CPHT-RR-017.0417, DFTT-2007-5, LPT-ORSAY-07-23",
    doi = "10.1016/j.physletb.2007.06.031",
    journal = "Phys. Lett. B",
    volume = "651",
    pages = "239--245",
    year = "2007"
}

@article{Vafa:2005ui,
    author = "Vafa, Cumrun",
    title = "{The String landscape and the swampland}",
    eprint = "hep-th/0509212",
    archivePrefix = "arXiv",
    reportNumber = "HUTP-05-A043",
    month = "9",
    year = "2005"
}

@article{Palti:2019pca,
    author = "Palti, Eran",
    title = "{The Swampland: Introduction and Review}",
    eprint = "1903.06239",
    archivePrefix = "arXiv",
    primaryClass = "hep-th",
    reportNumber = "MPP-2019-53",
    doi = "10.1002/prop.201900037",
    journal = "Fortsch. Phys.",
    volume = "67",
    number = "6",
    pages = "1900037",
    year = "2019"
}

@article{vanBeest:2021lhn,
    author = "van Beest, Marieke and Calder\'on-Infante, Jos\'e and Mirfendereski, Delaram and Valenzuela, Irene",
    title = "{Lectures on the Swampland Program in String Compactifications}",
    eprint = "2102.01111",
    archivePrefix = "arXiv",
    primaryClass = "hep-th",
    month = "2",
    year = "2021"
}

@article{Grana:2021zvf,
    author = "Gra\~na, Mariana and Herr\'aez, Alvaro",
    title = "{The Swampland Conjectures: A Bridge from Quantum Gravity to Particle Physics}",
    eprint = "2107.00087",
    archivePrefix = "arXiv",
    primaryClass = "hep-th",
    doi = "10.3390/universe7080273",
    journal = "Universe",
    volume = "7",
    number = "8",
    pages = "273",
    year = "2021"
}

@article{Ooguri:2006in,
    author = "Ooguri, Hirosi and Vafa, Cumrun",
    title = "{On the Geometry of the String Landscape and the Swampland}",
    eprint = "hep-th/0605264",
    archivePrefix = "arXiv",
    reportNumber = "CALT-68-2600, HUTP-06-A017",
    doi = "10.1016/j.nuclphysb.2006.10.033",
    journal = "Nucl. Phys. B",
    volume = "766",
    pages = "21--33",
    year = "2007"
}

@article{Baume:2020dqd,
    author = "Baume, Florent and Calder\'on Infante, Jos\'e",
    title = "{Tackling the SDC in AdS with CFTs}",
    eprint = "2011.03583",
    archivePrefix = "arXiv",
    primaryClass = "hep-th",
    reportNumber = "IFT-UAM/CSIC-20-142",
    doi = "10.1007/JHEP08(2021)057",
    journal = "JHEP",
    volume = "08",
    pages = "057",
    year = "2021"
}

@article{Perlmutter:2020buo,
    author = "Perlmutter, Eric and Rastelli, Leonardo and Vafa, Cumrun and Valenzuela, Irene",
    title = "{A CFT distance conjecture}",
    eprint = "2011.10040",
    archivePrefix = "arXiv",
    primaryClass = "hep-th",
    doi = "10.1007/JHEP10(2021)070",
    journal = "JHEP",
    volume = "10",
    pages = "070",
    year = "2021"
}

@article{Montero:2020icj,
    author = "Montero, Miguel and Vafa, Cumrun",
    title = "{Cobordism Conjecture, Anomalies, and the String Lamppost Principle}",
    eprint = "2008.11729",
    archivePrefix = "arXiv",
    primaryClass = "hep-th",
    doi = "10.1007/JHEP01(2021)063",
    journal = "JHEP",
    volume = "01",
    pages = "063",
    year = "2021"
}

@article{Hamada:2021bbz,
    author = "Hamada, Yuta and Vafa, Cumrun",
    title = "{8d supergravity, reconstruction of internal geometry and the Swampland}",
    eprint = "2104.05724",
    archivePrefix = "arXiv",
    primaryClass = "hep-th",
    doi = "10.1007/JHEP06(2021)178",
    journal = "JHEP",
    volume = "06",
    pages = "178",
    year = "2021"
}

@article{Tarazi:2021duw,
    author = "Tarazi, Houri-Christina and Vafa, Cumrun",
    title = "{On The Finiteness of 6d Supergravity Landscape}",
    eprint = "2106.10839",
    archivePrefix = "arXiv",
    primaryClass = "hep-th",
    month = "6",
    year = "2021"
}

@article{Bedroya:2021fbu,
    author = "Bedroya, Alek and Hamada, Yuta and Montero, Miguel and Vafa, Cumrun",
    title = "{Compactness of brane moduli and the String Lamppost Principle in d \ensuremath{>} 6}",
    eprint = "2110.10157",
    archivePrefix = "arXiv",
    primaryClass = "hep-th",
    doi = "10.1007/JHEP02(2022)082",
    journal = "JHEP",
    volume = "02",
    pages = "082",
    year = "2022"
}

@article{Cribiori:2021gbf,
    author = "Cribiori, Niccol\`o and Lust, Dieter and Scalisi, Marco",
    title = "{The gravitino and the swampland}",
    eprint = "2104.08288",
    archivePrefix = "arXiv",
    primaryClass = "hep-th",
    reportNumber = "LMU-ASC 10/21, MPP-2021-62",
    doi = "10.1007/JHEP06(2021)071",
    journal = "JHEP",
    volume = "06",
    pages = "071",
    year = "2021"
}

@article{Castellano:2021yye,
    author = "Castellano, Alberto and Font, Anamar\'\i{}a and Herraez, Alvaro and Ib\'a\~nez, Luis E.",
    title = "{A gravitino distance conjecture}",
    eprint = "2104.10181",
    archivePrefix = "arXiv",
    primaryClass = "hep-th",
    doi = "10.1007/JHEP08(2021)092",
    journal = "JHEP",
    volume = "08",
    pages = "092",
    year = "2021"
}

@article{DallAgata:2021nnr,
    author = "Dall'Agata, Gianguido and Emelin, Maxim and Farakos, Fotis and Morittu, Matteo",
    title = "{The unbearable lightness of charged gravitini}",
    eprint = "2108.04254",
    archivePrefix = "arXiv",
    primaryClass = "hep-th",
    doi = "10.1007/JHEP10(2021)076",
    journal = "JHEP",
    volume = "10",
    pages = "076",
    year = "2021"
}

@article{Antoniadis:1999xk,
    author = "Antoniadis, Ignatios and Dudas, E. and Sagnotti, A.",
    title = "{Brane supersymmetry breaking}",
    eprint = "hep-th/9908023",
    archivePrefix = "arXiv",
    reportNumber = "CPHT-S727-0799, LPT-ORSAY-99-60, ROM2F-99-23",
    doi = "10.1016/S0370-2693(99)01023-0",
    journal = "Phys. Lett. B",
    volume = "464",
    pages = "38--45",
    year = "1999"
}

@article{Angelantonj:1999jh,
    author = "Angelantonj, Carlo",
    title = "{Comments on open string orbifolds with a nonvanishing B(ab)}",
    eprint = "hep-th/9908064",
    archivePrefix = "arXiv",
    reportNumber = "CPHT-S718-0599, LPTENS-99-27, CPHT-718.0599",
    doi = "10.1016/S0550-3213(99)00662-8",
    journal = "Nucl. Phys. B",
    volume = "566",
    pages = "126--150",
    year = "2000"
}

@article{Aldazabal:1999jr,
    author = "Aldazabal, G. and Uranga, A. M.",
    title = "{Tachyon free nonsupersymmetric type IIB orientifolds via Brane - anti-brane systems}",
    eprint = "hep-th/9908072",
    archivePrefix = "arXiv",
    reportNumber = "CAB-IB-2911299, IASSNS-HEP-99-79",
    doi = "10.1088/1126-6708/1999/10/024",
    journal = "JHEP",
    volume = "10",
    pages = "024",
    year = "1999"
}

@article{Angelantonj:1999ms,
    author = "Angelantonj, C. and Antoniadis, Ignatios and D'Appollonio, G. and Dudas, E. and Sagnotti, A.",
    title = "{Type I vacua with brane supersymmetry breaking}",
    eprint = "hep-th/9911081",
    archivePrefix = "arXiv",
    reportNumber = "LPTENS-99-38, CPTH-S743-1099, NSF-ITP-99-127, DFF-347-10-99, ROM2F-99-40, CPTH-S743.1099, LPT-ORSAY-99-80",
    doi = "10.1016/S0550-3213(00)00052-3",
    journal = "Nucl. Phys. B",
    volume = "572",
    pages = "36--70",
    year = "2000"
}

@article{Coudarchet:2021qwc,
    author = "Coudarchet, Thibaut and Dudas, Emilian and Partouche, Herv\'e",
    title = "{Geometry of orientifold vacua and supersymmetry breaking}",
    eprint = "2105.06913",
    archivePrefix = "arXiv",
    primaryClass = "hep-th",
    reportNumber = "CPHT-RR026.032021",
    doi = "10.1007/JHEP07(2021)104",
    journal = "JHEP",
    volume = "07",
    pages = "104",
    year = "2021"
}

@article{AlvarezGaume:1986jb,
    author = "Alvarez-Gaume, Luis and Ginsparg, Paul H. and Moore, Gregory W. and Vafa, C.",
    title = "{An O(16) x O(16) Heterotic String}",
    reportNumber = "HUTP-86/A013",
    doi = "10.1016/0370-2693(86)91524-8",
    journal = "Phys. Lett. B",
    volume = "171",
    pages = "155--162",
    year = "1986"
}

@article{Dixon:1986iz,
    author = "Dixon, Lance J. and Harvey, Jeffrey A.",
    editor = "Schellekens, B.",
    title = "{String Theories in Ten-Dimensions Without Space-Time Supersymmetry}",
    reportNumber = "PRINT-86-0244 (PRINCETON)",
    doi = "10.1016/0550-3213(86)90619-X",
    journal = "Nucl. Phys. B",
    volume = "274",
    pages = "93--105",
    year = "1986"
}

@inproceedings{Sagnotti:1995ga,
    author = "Sagnotti, Augusto",
    title = "{Some properties of open string theories}",
    booktitle = "{International Workshop on Supersymmetry and Unification of Fundamental Interactions (SUSY 95)}",
    eprint = "hep-th/9509080",
    archivePrefix = "arXiv",
    reportNumber = "ROM2F-95-18",
    pages = "473--484",
    month = "9",
    year = "1995"
}

@article{Sagnotti:1996qj,
    author = "Sagnotti, Augusto",
    editor = "Lust, D. and Otto, H. J. and Weigt, G.",
    title = "{Surprises in open string perturbation theory}",
    eprint = "hep-th/9702093",
    archivePrefix = "arXiv",
    reportNumber = "ROM2F-97-4",
    doi = "10.1016/S0920-5632(97)00344-7",
    journal = "Nucl. Phys. B Proc. Suppl.",
    volume = "56",
    pages = "332--343",
    year = "1997"
}

@article{Sugimoto:1999tx,
    author = "Sugimoto, Shigeki",
    title = "{Anomaly cancellations in type I D-9 - anti-D-9 system and the USp(32) string theory}",
    eprint = "hep-th/9905159",
    archivePrefix = "arXiv",
    reportNumber = "YITP-99-25",
    doi = "10.1143/PTP.102.685",
    journal = "Prog. Theor. Phys.",
    volume = "102",
    pages = "685--699",
    year = "1999"
}

@article{Mourad:2020cjq,
    author = "Mourad, J. and Sagnotti, A.",
    title = "{On boundaries, charges and Fermi fields}",
    eprint = "2002.05372",
    archivePrefix = "arXiv",
    primaryClass = "hep-th",
    doi = "10.1016/j.physletb.2020.135368",
    journal = "Phys. Lett. B",
    volume = "804",
    pages = "135368",
    year = "2020"
}

@article{Gubser:2001zr,
    author = "Gubser, Steven S. and Mitra, Indrajit",
    title = "{Some interesting violations of the Breitenlohner-Freedman bound}",
    eprint = "hep-th/0108239",
    archivePrefix = "arXiv",
    reportNumber = "CALT-68-2345, CITUSC-01-030, PUPT-2005",
    doi = "10.1088/1126-6708/2002/07/044",
    journal = "JHEP",
    volume = "07",
    pages = "044",
    year = "2002"
}

@article{Mourad:2016xbk,
    author = "Mourad, J. and Sagnotti, A.",
    title = "{$AdS$ Vacua from Dilaton Tadpoles and Form Fluxes}",
    eprint = "1612.08566",
    archivePrefix = "arXiv",
    primaryClass = "hep-th",
    doi = "10.1016/j.physletb.2017.02.053",
    journal = "Phys. Lett. B",
    volume = "768",
    pages = "92--96",
    year = "2017"
}

@article{Mourad:2017rrl,
    author = "Mourad, J. and Sagnotti, A.",
    title = "{An Update on Brane Supersymmetry Breaking}",
    eprint = "1711.11494",
    archivePrefix = "arXiv",
    primaryClass = "hep-th",
    month = "11",
    year = "2017"
}

@article{Fre:2013vza,
    author = "Fr\'e, P. and Sagnotti, A. and Sorin, A. S.",
    title = "{Integrable Scalar Cosmologies I. Foundations and links with String Theory}",
    eprint = "1307.1910",
    archivePrefix = "arXiv",
    primaryClass = "hep-th",
    doi = "10.1016/j.nuclphysb.2013.10.015",
    journal = "Nucl. Phys. B",
    volume = "877",
    pages = "1028--1106",
    year = "2013"
}

@phdthesis{Basile:2020xwi,
    author = "Basile, Ivano",
    title = "{On String Vacua without Supersymmetry: brane dynamics, bubbles and holography}",
    eprint = "2010.00628",
    archivePrefix = "arXiv",
    primaryClass = "hep-th",
    school = "Pisa, Scuola Normale Superiore",
    year = "2020"
}

@article{Basile:2021vxh,
    author = "Basile, Ivano",
    title = "{Supersymmetry breaking and stability in string vacua: Brane dynamics, bubbles and the swampland}",
    eprint = "2107.02814",
    archivePrefix = "arXiv",
    primaryClass = "hep-th",
    doi = "10.1007/s40766-021-00024-9",
    journal = "Riv. Nuovo Cim.",
    volume = "44",
    number = "10",
    pages = "499--596",
    year = "2021"
}

@article{Basile:2022zee,
    author = "Basile, Ivano",
    title = "{Emergent strings at infinite distance with broken supersymmetry}",
    eprint = "2201.08851",
    archivePrefix = "arXiv",
    primaryClass = "hep-th",
    month = "1",
    year = "2022"
}

@article{Mourad:2021lma,
    author = "Sagnotti, Augusto and Mourad, Jihad",
    title = "{String (In)Stability Issues with Broken Supersymmetry}",
    eprint = "2107.04064",
    archivePrefix = "arXiv",
    primaryClass = "hep-th",
    doi = "10.31526/lhep.2021.219",
    journal = "LHEP",
    volume = "2021",
    pages = "219",
    year = "2021"
}

@article{Pelliconi:2021eak,
    author = "Pelliconi, P. and Sagnotti, A.",
    title = "{Integrable Models and Supersymmetry Breaking}",
    eprint = "2102.06184",
    archivePrefix = "arXiv",
    primaryClass = "hep-th",
    doi = "10.1016/j.nuclphysb.2021.115363",
    journal = "Nucl. Phys. B",
    volume = "965",
    pages = "115363",
    year = "2021"
}

@article{Dudas:2000nv,
    author = "Dudas, E. and Mourad, J.",
    title = "{Consistent gravitino couplings in nonsupersymmetric strings}",
    eprint = "hep-th/0012071",
    archivePrefix = "arXiv",
    reportNumber = "LPT-ORSAY-00-128",
    doi = "10.1016/S0370-2693(01)00777-8",
    journal = "Phys. Lett. B",
    volume = "514",
    pages = "173--182",
    year = "2001"
}

@article{Pradisi:2001yv,
    author = "Pradisi, Gianfranco and Riccioni, Fabio",
    title = "{Geometric couplings and brane supersymmetry breaking}",
    eprint = "hep-th/0107090",
    archivePrefix = "arXiv",
    reportNumber = "ROM2F-01-24",
    doi = "10.1016/S0550-3213(01)00441-2",
    journal = "Nucl. Phys. B",
    volume = "615",
    pages = "33--60",
    year = "2001"
}

@article{Dudas:2000ff,
    author = "Dudas, E. and Mourad, J.",
    title = "{Brane solutions in strings with broken supersymmetry and dilaton tadpoles}",
    eprint = "hep-th/0004165",
    archivePrefix = "arXiv",
    reportNumber = "LPT-ORSAY-00-43, LPTM-00-25, LPT-00-43",
    doi = "10.1016/S0370-2693(00)00734-6",
    journal = "Phys. Lett. B",
    volume = "486",
    pages = "172--178",
    year = "2000"
}

@article{Dudas:2001wd,
    author = "Dudas, E. and Mourad, J. and Sagnotti, A.",
    title = "{Charged and uncharged D-branes in various string theories}",
    eprint = "hep-th/0107081",
    archivePrefix = "arXiv",
    reportNumber = "LPT-ORSAY-01-56, ROM2F-01-18",
    doi = "10.1016/S0550-3213(01)00552-1",
    journal = "Nucl. Phys. B",
    volume = "620",
    pages = "109--151",
    year = "2002"
}

@article{Antonelli:2019nar,
    author = "Antonelli, Riccardo and Basile, Ivano",
    title = "{Brane annihilation in non-supersymmetric strings}",
    eprint = "1908.04352",
    archivePrefix = "arXiv",
    primaryClass = "hep-th",
    doi = "10.1007/JHEP11(2019)021",
    journal = "JHEP",
    volume = "11",
    pages = "021",
    year = "2019"
}

@article{Buratti:2021fiv,
    author = "Buratti, Ginevra and Calder\'on-Infante, Jos\'e and Delgado, Matilda and Uranga, Angel M.",
    title = "{Dynamical Cobordism and Swampland Distance Conjectures}",
    eprint = "2107.09098",
    archivePrefix = "arXiv",
    primaryClass = "hep-th",
    doi = "10.1007/JHEP10(2021)037",
    journal = "JHEP",
    volume = "10",
    pages = "037",
    year = "2021"
}

@article{Buratti:2021yia,
    author = "Buratti, Ginevra and Delgado, Matilda and Uranga, Angel M.",
    title = "{Dynamical tadpoles, stringy cobordism, and the SM from spontaneous compactification}",
    eprint = "2104.02091",
    archivePrefix = "arXiv",
    primaryClass = "hep-th",
    doi = "10.1007/JHEP06(2021)170",
    journal = "JHEP",
    volume = "06",
    pages = "170",
    year = "2021"
}

@article{Angius:2022aeq,
    author = "Angius, Roberta and Calder\'on-Infante, Jos\'e and Delgado, Matilda and Huertas, Jes\'us and Uranga, Angel M.",
    title = "{At the End of the World: Local Dynamical Cobordism}",
    eprint = "2203.11240",
    archivePrefix = "arXiv",
    primaryClass = "hep-th",
    reportNumber = "IFT-UAM/CSIC-22-31",
    month = "3",
    year = "2022"
}

@article{Angius:2022mgh,
    author = "Angius, Roberta and Delgado, Matilda and Uranga, Angel M.",
    title = "{Dynamical Cobordism and the Beginning of Time: Supercritical Strings and Tachyon Condensation}",
    eprint = "2207.13108",
    archivePrefix = "arXiv",
    primaryClass = "hep-th",
    month = "7",
    year = "2022"
}

@article{Garcia-Etxebarria:2018ajm,
    author = "Garc\'\i{}a-Etxebarria, I\~naki and Montero, Miguel",
    title = "{Dai-Freed anomalies in particle physics}",
    eprint = "1808.00009",
    archivePrefix = "arXiv",
    primaryClass = "hep-th",
    reportNumber = "MPP-2018-188",
    doi = "10.1007/JHEP08(2019)003",
    journal = "JHEP",
    volume = "08",
    pages = "003",
    year = "2019"
}

@article{Debray:2021vob,
    author = "Debray, Arun and Dierigl, Markus and Heckman, Jonathan J. and Montero, Miguel",
    title = "{The anomaly that was not meant IIB}",
    eprint = "2107.14227",
    archivePrefix = "arXiv",
    primaryClass = "hep-th",
    reportNumber = "LMU-ASC 24/21",
    doi = "10.1002/prop.202100168",
    month = "7",
    year = "2021"
}

@article{Blumenhagen:2000dc,
    author = "Blumenhagen, Ralph and Font, Anamaria",
    title = "{Dilaton tadpoles, warped geometries and large extra dimensions for nonsupersymmetric strings}",
    eprint = "hep-th/0011269",
    archivePrefix = "arXiv",
    reportNumber = "HUB-EP-00-55",
    doi = "10.1016/S0550-3213(01)00028-1",
    journal = "Nucl. Phys. B",
    volume = "599",
    pages = "241--254",
    year = "2001"
}

@article{Dudas:2002dg,
    author = "Dudas, E. and Mourad, J. and Timirgaziu, Cristina",
    title = "{Time and space dependent backgrounds from nonsupersymmetric strings}",
    eprint = "hep-th/0209176",
    archivePrefix = "arXiv",
    reportNumber = "CPHT-RR-068-0902, LPT-ORSAY-02-136",
    doi = "10.1016/S0550-3213(03)00248-7",
    journal = "Nucl. Phys. B",
    volume = "660",
    pages = "3--24",
    year = "2003"
}

@article{Dudas:2010gi,
    author = "Dudas, E. and Kitazawa, N. and Sagnotti, A.",
    title = "{On Climbing Scalars in String Theory}",
    eprint = "1009.0874",
    archivePrefix = "arXiv",
    primaryClass = "hep-th",
    reportNumber = "CPHT-RR078.0910, LPT-ORSAY-10-65",
    doi = "10.1016/j.physletb.2010.09.040",
    journal = "Phys. Lett. B",
    volume = "694",
    pages = "80--88",
    year = "2011"
}

@article{DeWolfe:2005uu,
    author = "DeWolfe, Oliver and Giryavets, Alexander and Kachru, Shamit and Taylor, Washington",
    title = "{Type IIA moduli stabilization}",
    eprint = "hep-th/0505160",
    archivePrefix = "arXiv",
    reportNumber = "MIT-CTP-3640, PUPT-2161, SU-ITP-05-16, SLAC-PUB-11153",
    doi = "10.1088/1126-6708/2005/07/066",
    journal = "JHEP",
    volume = "07",
    pages = "066",
    year = "2005"
}

@article{Tsimpis:2012tu,
    author = "Tsimpis, Dimitrios",
    title = "{Supersymmetric AdS vacua and separation of scales}",
    eprint = "1206.5900",
    archivePrefix = "arXiv",
    primaryClass = "hep-th",
    doi = "10.1007/JHEP08(2012)142",
    journal = "JHEP",
    volume = "08",
    pages = "142",
    year = "2012"
}

@article{Gautason:2015tig,
    author = "Gautason, F. F. and Schillo, M. and Van Riet, T. and Williams, M.",
    title = "{Remarks on scale separation in flux vacua}",
    eprint = "1512.00457",
    archivePrefix = "arXiv",
    primaryClass = "hep-th",
    doi = "10.1007/JHEP03(2016)061",
    journal = "JHEP",
    volume = "03",
    pages = "061",
    year = "2016"
}

@article{Lust:2019zwm,
    author = {L\"ust, Dieter and Palti, Eran and Vafa, Cumrun},
    title = "{AdS and the Swampland}",
    eprint = "1906.05225",
    archivePrefix = "arXiv",
    primaryClass = "hep-th",
    doi = "10.1016/j.physletb.2019.134867",
    journal = "Phys. Lett. B",
    volume = "797",
    pages = "134867",
    year = "2019"
}

@article{Buratti:2020kda,
    author = "Buratti, Ginevra and Calderon, Jose and Mininno, Alessandro and Uranga, Angel M.",
    title = "{Discrete Symmetries, Weak Coupling Conjecture and Scale Separation in AdS Vacua}",
    eprint = "2003.09740",
    archivePrefix = "arXiv",
    primaryClass = "hep-th",
    reportNumber = "IFT-UAM/CSIC-20-46",
    doi = "10.1007/JHEP06(2020)083",
    journal = "JHEP",
    volume = "06",
    pages = "083",
    year = "2020"
}

@article{Lust:2020npd,
    author = {L\"ust, Dieter and Tsimpis, Dimitrios},
    title = "{AdS$_{2}$ type-IIA solutions and scale separation}",
    eprint = "2004.07582",
    archivePrefix = "arXiv",
    primaryClass = "hep-th",
    reportNumber = "LMU-ASC 15/20, MPP-2020-50",
    doi = "10.1007/JHEP07(2020)060",
    journal = "JHEP",
    volume = "07",
    pages = "060",
    year = "2020"
}

@article{Junghans:2020acz,
    author = "Junghans, Daniel",
    title = "{O-Plane Backreaction and Scale Separation in Type IIA Flux Vacua}",
    eprint = "2003.06274",
    archivePrefix = "arXiv",
    primaryClass = "hep-th",
    doi = "10.1002/prop.202000040",
    journal = "Fortsch. Phys.",
    volume = "68",
    number = "6",
    pages = "2000040",
    year = "2020"
}

@article{Marchesano:2020qvg,
    author = "Marchesano, Fernando and Palti, Eran and Quirant, Joan and Tomasiello, Alessandro",
    title = "{On supersymmetric AdS$_{4}$ orientifold vacua}",
    eprint = "2003.13578",
    archivePrefix = "arXiv",
    primaryClass = "hep-th",
    reportNumber = "IFT-UAM/CSIC-20-51",
    doi = "10.1007/JHEP08(2020)087",
    journal = "JHEP",
    volume = "08",
    pages = "087",
    year = "2020"
}

@article{Marchesano:2020uqz,
    author = "Marchesano, Fernando and Prieto, David and Quirant, Joan and Shukla, Pramod",
    title = "{Systematics of Type IIA moduli stabilisation}",
    eprint = "2007.00672",
    archivePrefix = "arXiv",
    primaryClass = "hep-th",
    reportNumber = "IFT-UAM/CSIC-20-95",
    doi = "10.1007/JHEP11(2020)113",
    journal = "JHEP",
    volume = "11",
    pages = "113",
    year = "2020"
}

@article{DeLuca:2021mcj,
    author = "De Luca, G. Bruno and Tomasiello, Alessandro",
    title = "{Leaps and bounds towards scale separation}",
    eprint = "2104.12773",
    archivePrefix = "arXiv",
    primaryClass = "hep-th",
    doi = "10.1007/JHEP12(2021)086",
    journal = "JHEP",
    volume = "12",
    pages = "086",
    year = "2021"
}

@article{DeLuca:2021ojx,
    author = "De Luca, G. Bruno and De Ponti, Nicol\`o and Mondino, Andrea and Tomasiello, Alessandro",
    title = "{Cheeger bounds on spin-two fields}",
    eprint = "2109.11560",
    archivePrefix = "arXiv",
    primaryClass = "hep-th",
    doi = "10.1007/JHEP12(2021)217",
    journal = "JHEP",
    volume = "12",
    pages = "217",
    year = "2021"
}

@article{Cribiori:2021djm,
    author = "Cribiori, Niccol\`o and Junghans, Daniel and Van Hemelryck, Vincent and Van Riet, Thomas and Wrase, Timm",
    title = "{Scale-separated AdS4 vacua of IIA orientifolds and M-theory}",
    eprint = "2107.00019",
    archivePrefix = "arXiv",
    primaryClass = "hep-th",
    reportNumber = "UUITP-29/21",
    doi = "10.1103/PhysRevD.104.126014",
    journal = "Phys. Rev. D",
    volume = "104",
    number = "12",
    pages = "126014",
    year = "2021"
}

@article{Emelin:2022cac,
    author = "Emelin, Maxim and Farakos, Fotis and Tringas, George",
    title = "{O6-plane backreaction on scale-separated Type IIA AdS$_3$ vacua}",
    eprint = "2202.13431",
    archivePrefix = "arXiv",
    primaryClass = "hep-th",
    month = "2",
    year = "2022"
}

@article{Andriot:2022yyj,
    author = "Andriot, David and Horer, Ludwig and Marconnet, Paul",
    title = "{Exploring the landscape of (anti-) de Sitter and Minkowski solutions: group manifolds, stability and scale separation}",
    eprint = "2204.05327",
    archivePrefix = "arXiv",
    primaryClass = "hep-th",
    month = "4",
    year = "2022"
}

@article{Cribiori:2022trc,
    author = "Cribiori, Niccol\`o and Dall'Agata, Gianguido",
    title = "{Weak gravity versus scale separation}",
    eprint = "2203.05559",
    archivePrefix = "arXiv",
    primaryClass = "hep-th",
    reportNumber = "MPP-2022-26",
    month = "3",
    year = "2022"
}

@article{Apers:2022zjx,
    author = "Apers, Fien and Montero, Miguel and Van Riet, Thomas and Wrase, Timm",
    title = "{Comments on classical AdS flux vacua with scale separation}",
    eprint = "2202.00682",
    archivePrefix = "arXiv",
    primaryClass = "hep-th",
    month = "2",
    year = "2022"
}

@article{Tsimpis:2022orc,
    author = "Tsimpis, Dimitrios",
    title = "{Relative scale separation in orbifolds of S$^{2}$ and S$^{5}$}",
    eprint = "2201.10916",
    archivePrefix = "arXiv",
    primaryClass = "hep-th",
    doi = "10.1007/JHEP03(2022)169",
    journal = "JHEP",
    volume = "03",
    pages = "169",
    year = "2022"
}

@article{VanHemelryck:2022ynr,
    author = "Van Hemelryck, Vincent",
    title = "{Scale-separated AdS$_3$ vacua from $G_2$-orientifolds using pure spinors}",
    eprint = "2207.14311",
    archivePrefix = "arXiv",
    primaryClass = "hep-th",
    month = "7",
    year = "2022"
}

@article{Ooguri:2016pdq,
    author = "Ooguri, Hirosi and Vafa, Cumrun",
    title = "{Non-supersymmetric AdS and the Swampland}",
    eprint = "1610.01533",
    archivePrefix = "arXiv",
    primaryClass = "hep-th",
    reportNumber = "CALT-TH-2016-027, IPMU16-0139",
    doi = "10.4310/ATMP.2017.v21.n7.a8",
    journal = "Adv. Theor. Math. Phys.",
    volume = "21",
    pages = "1787--1801",
    year = "2017"
}

@article{Freivogel:2016qwc,
    author = "Freivogel, Ben and Kleban, Matthew",
    title = "{Vacua Morghulis}",
    eprint = "1610.04564",
    archivePrefix = "arXiv",
    primaryClass = "hep-th",
    month = "10",
    year = "2016"
}

@article{Basile:2018irz,
    author = "Basile, I. and Mourad, J. and Sagnotti, A.",
    title = "{On Classical Stability with Broken Supersymmetry}",
    eprint = "1811.11448",
    archivePrefix = "arXiv",
    primaryClass = "hep-th",
    doi = "10.1007/JHEP01(2019)174",
    journal = "JHEP",
    volume = "01",
    pages = "174",
    year = "2019"
}

@article{Mourad:2022loy,
    author = "Mourad, J. and Sagnotti, A.",
    title = "{A 4D IIB Flux Vacuum and Supersymmetry Breaking. I. Fermionic Spectrum}",
    eprint = "2206.03340",
    archivePrefix = "arXiv",
    primaryClass = "hep-th",
    month = "6",
    year = "2022"
}

@article{Dibitetto:2020csn,
    author = "Dibitetto, Giuseppe and Petri, Nicol\`o and Schillo, Marjorie",
    title = "{Nothing really matters}",
    eprint = "2002.01764",
    archivePrefix = "arXiv",
    primaryClass = "hep-th",
    doi = "10.1007/JHEP08(2020)040",
    journal = "JHEP",
    volume = "08",
    pages = "040",
    year = "2020"
}

@article{GarciaEtxebarria:2020xsr,
    author = "Garc\'\i{}a Etxebarria, I\~naki and Montero, Miguel and Sousa, Kepa and Valenzuela, Irene",
    title = "{Nothing is certain in string compactifications}",
    eprint = "2005.06494",
    archivePrefix = "arXiv",
    primaryClass = "hep-th",
    doi = "10.1007/JHEP12(2020)032",
    journal = "JHEP",
    volume = "12",
    pages = "032",
    year = "2020"
}

@article{Guarino:2020flh,
    author = "Guarino, Adolfo and Malek, Emanuel and Samtleben, Henning",
    title = "{Stable Nonsupersymmetric Anti\textendash{}de Sitter Vacua of Massive IIA Supergravity}",
    eprint = "2011.06600",
    archivePrefix = "arXiv",
    primaryClass = "hep-th",
    reportNumber = "HU-EP-20/34",
    doi = "10.1103/PhysRevLett.126.061601",
    journal = "Phys. Rev. Lett.",
    volume = "126",
    number = "6",
    pages = "061601",
    year = "2021"
}

@article{Bomans:2021ara,
    author = "Bomans, Pieter and Cassani, Davide and Dibitetto, Giuseppe and Petri, Nicolo",
    title = "{Bubble instability of mIIA on $\mathrm{AdS}_4\times S^6$}",
    eprint = "2110.08276",
    archivePrefix = "arXiv",
    primaryClass = "hep-th",
    month = "10",
    year = "2021"
}

@article{Guarino:2021hrc,
    author = "Guarino, Adolfo and Sterckx, Colin",
    title = "{Flat deformations of type IIB S-folds}",
    eprint = "2109.06032",
    archivePrefix = "arXiv",
    primaryClass = "hep-th",
    doi = "10.1007/JHEP11(2021)171",
    journal = "JHEP",
    volume = "11",
    pages = "171",
    year = "2021"
}

@article{Giambrone:2021wsm,
    author = "Giambrone, Alfredo and Guarino, Adolfo and Malek, Emanuel and Samtleben, Henning and Sterckx, Colin and Trigiante, Mario",
    title = "{Holographic evidence for nonsupersymmetric conformal manifolds}",
    eprint = "2112.11966",
    archivePrefix = "arXiv",
    primaryClass = "hep-th",
    doi = "10.1103/PhysRevD.105.066018",
    journal = "Phys. Rev. D",
    volume = "105",
    number = "6",
    pages = "066018",
    year = "2022"
}

@inproceedings{Guarino:2022tlw,
    author = "Guarino, Adolfo and Sterckx, Colin",
    title = "{Type IIB S-folds: flat deformations, holography and stability}",
    booktitle = "{21st Hellenic School and Workshops on Elementary Particle Physics and Gravity}",
    eprint = "2204.09993",
    archivePrefix = "arXiv",
    primaryClass = "hep-th",
    month = "4",
    year = "2022"
}

@article{Dibitetto:2022rzy,
    author = "Dibitetto, Giuseppe and Petri, Nicol\`o",
    title = "{Searching for Coleman-de Luccia bubbles in AdS compactifications}",
    eprint = "2207.02172",
    archivePrefix = "arXiv",
    primaryClass = "hep-th",
    month = "7",
    year = "2022"
}

@article{Montero:2022prj,
    author = "Montero, Miguel and Vafa, Cumrun and Valenzuela, Irene",
    title = "{The Dark Dimension and the Swampland}",
    eprint = "2205.12293",
    archivePrefix = "arXiv",
    primaryClass = "hep-th",
    month = "5",
    year = "2022"
}

@article{Anchordoqui:2022txe,
    author = "Anchordoqui, Luis and Antoniadis, Ignatios and Lust, Dieter",
    title = "{The Dark Dimension, the Swampland, and the Dark Matter Fraction Composed of Primordial Black Holes}",
    eprint = "2206.07071",
    archivePrefix = "arXiv",
    primaryClass = "hep-th",
    reportNumber = "MPP-2022-60, LMU-ASC 24/22",
    month = "6",
    year = "2022"
}

@article{Blumenhagen:2022zzw,
    author = "Blumenhagen, Ralph and Brinkmann, Max and Makridou, Andriana",
    title = "{The Dark Dimension in a Warped Throat}",
    eprint = "2208.01057",
    archivePrefix = "arXiv",
    primaryClass = "hep-th",
    reportNumber = "MPP-2022-94",
    month = "8",
    year = "2022"
}

@article{McNamara:2019rup,
    author = "McNamara, Jacob and Vafa, Cumrun",
    title = "{Cobordism Classes and the Swampland}",
    eprint = "1909.10355",
    archivePrefix = "arXiv",
    primaryClass = "hep-th",
    month = "9",
    year = "2019"
}

@article{Blumenhagen:2021nmi,
    author = "Blumenhagen, Ralph and Cribiori, Niccol\`o",
    title = "{Open-Closed Correspondence of K-theory and Cobordism}",
    eprint = "2112.07678",
    archivePrefix = "arXiv",
    primaryClass = "hep-th",
    reportNumber = "MPP-2021-201",
    month = "12",
    year = "2021"
}

@article{Andriot:2022mri,
    author = "Andriot, David and Carqueville, Nils and Cribiori, Niccol\`o",
    title = "{Looking for structure in the cobordism conjecture}",
    eprint = "2204.00021",
    archivePrefix = "arXiv",
    primaryClass = "hep-th",
    reportNumber = "MPP-2022-11",
    month = "3",
    year = "2022"
}

@article{Blumenhagen:2022mqw,
    author = "Blumenhagen, Ralph and Cribiori, Niccol\`o and Kneissl, Christian and Makridou, Andriana",
    title = "{Dynamical Cobordism of a Domain Wall and its Companion Defect 7-brane}",
    eprint = "2205.09782",
    archivePrefix = "arXiv",
    primaryClass = "hep-th",
    reportNumber = "MPP-2022-57",
    month = "5",
    year = "2022"
}

@article{Raucci:2022jgw,
    author = "Raucci, Salvatore",
    title = "{On Codimension-one Vacua and String Theory}",
    eprint = "2206.06399",
    archivePrefix = "arXiv",
    primaryClass = "hep-th",
    month = "6",
    year = "2022"
}

@article{Blumenhagen:2022bvh,
    author = "Blumenhagen, Ralph and Cribiori, Niccol\`o and Kneissl, Christian and Makridou, Andriana",
    title = "{Dimensional Reduction of Cobordism and K-theory}",
    eprint = "2208.01656",
    archivePrefix = "arXiv",
    primaryClass = "hep-th",
    reportNumber = "MPP-2022-95",
    month = "8",
    year = "2022"
}

@inproceedings{Sagnotti:1987tw,
	title        = {{Open Strings and their Symmetry Groups}},
	author       = {Sagnotti, Augusto},
	year         = 1987,
	booktitle    = {{NATO Advanced Summer Institute on Nonperturbative Quantum Field Theory (Cargese Summer Institute) Cargese, France, July 16-30, 1987}},
	pages        = {521--528},
	eprint       = {hep-th/0208020},
	archiveprefix = {arXiv},
	primaryclass = {hep-th},
	reportnumber = {ROM2F-87-25},
	slaccitation = {%%CITATION = HEP-TH/0208020;%%}
}

@article{Pradisi:1988xd,
	title        = {{Open String Orbifolds}},
	author       = {Pradisi, Gianfranco and Sagnotti, Augusto},
	year         = 1989,
	journal      = {Phys. Lett.},
	volume       = {B216},
	pages        = {59--67},
	doi          = {10.1016/0370-2693(89)91369-5},
	reportnumber = {ROM2F-88-16},
	slaccitation = {%%CITATION = PHLTA,B216,59;%%}
}

@article{Horava:1989vt,
	title        = {{Strings on World Sheet Orbifolds}},
	author       = {Horava, Petr},
	year         = 1989,
	journal      = {Nucl. Phys.},
	volume       = {B327},
	pages        = {461--484},
	doi          = {10.1016/0550-3213(89)90279-4},
	reportnumber = {PRA-HEP-89/1},
	slaccitation = {%%CITATION = NUPHA,B327,461;%%}
}

@article{Horava:1989ga,
	title        = {{Background Duality of Open String Models}},
	author       = {Horava, Petr},
	year         = 1989,
	journal      = {Phys. Lett.},
	volume       = {B231},
	pages        = {251--257},
	doi          = {10.1016/0370-2693(89)90209-8},
	reportnumber = {PRA-HEP-89/4},
	slaccitation = {%%CITATION = PHLTA,B231,251;%%}
}

@article{Bianchi:1990yu,
	title        = {{On the systematics of open string theories}},
	author       = {Bianchi, Massimo and Sagnotti, Augusto},
	year         = 1990,
	journal      = {Phys. Lett.},
	volume       = {B247},
	pages        = {517--524},
	doi          = {10.1016/0370-2693(90)91894-H},
	reportnumber = {ROM2F-90-20},
	slaccitation = {%%CITATION = PHLTA,B247,517;%%}
}

@article{Bianchi:1990tb,
	title        = {{Twist symmetry and open string Wilson lines}},
	author       = {Bianchi, Massimo and Sagnotti, Augusto},
	year         = 1991,
	journal      = {Nucl. Phys.},
	volume       = {B361},
	pages        = {519--538},
	doi          = {10.1016/0550-3213(91)90271-X},
	reportnumber = {ROM2F-90-28},
	slaccitation = {%%CITATION = NUPHA,B361,519;%%}
}

@article{Bianchi:1991eu,
	title        = {{Toroidal compactification and symmetry breaking in open string theories}},
	author       = {Bianchi, M. and Pradisi, G. and Sagnotti, A.},
	year         = 1992,
	journal      = {Nucl. Phys.},
	volume       = {B376},
	pages        = {365--386},
	doi          = {10.1016/0550-3213(92)90129-Y},
	reportnumber = {ROM2F-91-15},
	slaccitation = {%%CITATION = NUPHA,B376,365;%%}
}

@article{Sagnotti:1992qw,
	title        = {{A Note on the Green-Schwarz mechanism in open string theories}},
	author       = {Sagnotti, Augusto},
	year         = 1992,
	journal      = {Phys. Lett.},
	volume       = {B294},
	pages        = {196--203},
	doi          = {10.1016/0370-2693(92)90682-T},
	eprint       = {hep-th/9210127},
	archiveprefix = {arXiv},
	primaryclass = {hep-th},
	reportnumber = {ROM2F-92-49},
	slaccitation = {%%CITATION = HEP-TH/9210127;%%}
}

@article{Lust:2022xoq,
    author = {L\"ust, Severin and Randall, Lisa},
    title = "{Effective Theory of Warped Compactifications and the Implications for KKLT}",
    eprint = "2206.04708",
    archivePrefix = "arXiv",
    primaryClass = "hep-th",
    month = "6",
    year = "2022"
}

@article{Leung:2022nhy,
    author = "Leung, Rahim and Stelle, K. S.",
    title = "{Supergravities on Branes}",
    eprint = "2205.13551",
    archivePrefix = "arXiv",
    primaryClass = "hep-th",
    reportNumber = "Imperial/TP/2022/KS/02",
    month = "5",
    year = "2022"
}

@article{Erickson:2021psj,
    author = "Erickson, C. W. and Leung, Rahim and Stelle, K. S.",
    title = "{Taxonomy of brane gravity localisations}",
    eprint = "2110.10688",
    archivePrefix = "arXiv",
    primaryClass = "hep-th",
    reportNumber = "Imperial/TP/21/KS/01",
    doi = "10.1007/JHEP01(2022)130",
    journal = "JHEP",
    volume = "01",
    pages = "130",
    year = "2022"
}

@article{Stelle:2020mmg,
    author = "Stelle, K. S.",
    title = "{Mass gaps and braneworlds}",
    eprint = "2004.00965",
    archivePrefix = "arXiv",
    primaryClass = "hep-th",
    doi = "10.1088/1751-8121/ab83ca",
    journal = "J. Phys. A",
    volume = "53",
    number = "20",
    pages = "204002",
    year = "2020"
}

@article{Basile:2020mpt,
    author = "Basile, Ivano and Lanza, Stefano",
    title = "{de Sitter in non-supersymmetric string theories: no-go theorems and brane-worlds}",
    eprint = "2007.13757",
    archivePrefix = "arXiv",
    primaryClass = "hep-th",
    doi = "10.1007/JHEP10(2020)108",
    journal = "JHEP",
    volume = "10",
    pages = "108",
    year = "2020"
}

@article{Banerjee:2019fzz,
    author = "Banerjee, Souvik and Danielsson, Ulf and Dibitetto, Giuseppe and Giri, Suvendu and Schillo, Marjorie",
    title = "{de Sitter Cosmology on an expanding bubble}",
    eprint = "1907.04268",
    archivePrefix = "arXiv",
    primaryClass = "hep-th",
    reportNumber = "UUITP-26/19",
    doi = "10.1007/JHEP10(2019)164",
    journal = "JHEP",
    volume = "10",
    pages = "164",
    year = "2019"
}

@article{Banerjee:2020wix,
    author = "Banerjee, Souvik and Danielsson, Ulf and Giri, Suvendu",
    title = "{Dark bubbles: decorating the wall}",
    eprint = "2001.07433",
    archivePrefix = "arXiv",
    primaryClass = "hep-th",
    reportNumber = "UUITP-1/20",
    doi = "10.1007/JHEP04(2020)085",
    journal = "JHEP",
    volume = "04",
    pages = "085",
    year = "2020"
}

@article{Banerjee:2020wov,
    author = "Banerjee, Souvik and Danielsson, Ulf and Giri, Suvendu",
    title = "{Bubble needs strings}",
    eprint = "2009.01597",
    archivePrefix = "arXiv",
    primaryClass = "hep-th",
    reportNumber = "UUITP-33/20",
    doi = "10.1007/JHEP03(2021)250",
    journal = "JHEP",
    volume = "21",
    pages = "250",
    year = "2020"
}

@article{Banerjee:2021qei,
    author = "Banerjee, Souvik and Danielsson, Ulf and Giri, Suvendu",
    title = "{Dark bubbles and black holes}",
    eprint = "2102.02164",
    archivePrefix = "arXiv",
    primaryClass = "hep-th",
    reportNumber = "UUITP-05/21",
    doi = "10.1007/JHEP09(2021)158",
    journal = "JHEP",
    volume = "09",
    pages = "158",
    year = "2021"
}

@article{Banerjee:2021yrb,
    author = "Banerjee, Souvik and Danielsson, Ulf and Giri, Suvendu",
    title = "{Curing with Hemlock: Escaping the swampland using instabilities from string theory}",
    eprint = "2103.17121",
    archivePrefix = "arXiv",
    primaryClass = "hep-th",
    reportNumber = "UUITP-15/21",
    doi = "10.1142/S0218271821420293",
    journal = "Int. J. Mod. Phys. D",
    volume = "30",
    number = "14",
    pages = "2142029",
    year = "2021"
}

@article{Danielsson:2021tyb,
    author = "Danielsson, U. H. and Panizo, D. and Tielemans, R. and Van Riet, T.",
    title = "{Higher-dimensional view on quantum cosmology}",
    eprint = "2105.03253",
    archivePrefix = "arXiv",
    primaryClass = "hep-th",
    reportNumber = "UUITP - 22/21",
    doi = "10.1103/PhysRevD.104.086015",
    journal = "Phys. Rev. D",
    volume = "104",
    number = "8",
    pages = "086015",
    year = "2021"
}

@article{Danielsson:2022fhd,
    author = "Danielsson, Ulf and Panizo, Daniel and Tielemans, Rob",
    title = "{Gravitational waves in dark bubble cosmology}",
    eprint = "2202.00545",
    archivePrefix = "arXiv",
    primaryClass = "hep-th",
    reportNumber = "UUITP - 03/22",
    doi = "10.1103/PhysRevD.106.024002",
    journal = "Phys. Rev. D",
    volume = "106",
    number = "2",
    pages = "024002",
    year = "2022"
}

@article{Chang:1979tg,
    author = "Chang, Shau-Jin and Weiss, Nathan",
    title = "{Instability of Constant {Yang-Mills} Fields}",
    reportNumber = "ILL-TH-79-3",
    doi = "10.1103/PhysRevD.20.869",
    journal = "Phys. Rev. D",
    volume = "20",
    pages = "869",
    year = "1979"
}

@article{Sikivie:1979bq,
    author = "Sikivie, P.",
    title = "{Instability of Abelian Field Configurations in {Yang-Mills} Theory}",
    reportNumber = "SLAC-PUB-2287",
    doi = "10.1103/PhysRevD.20.877",
    journal = "Phys. Rev. D",
    volume = "20",
    pages = "877",
    year = "1979"
}

@article{Raucci:2022bjw,
    author = "Raucci, Salvatore",
    title = "{On New Vacua of non-Supersymmetric Strings}",
    eprint = "2209.06537",
    archivePrefix = "arXiv",
    primaryClass = "hep-th",
    month = "9",
    year = "2022"
}

\end{document}